\documentclass{article}
\usepackage{graphicx}  
\usepackage{amsmath}   
\usepackage[compress,numbers,sort]{natbib}
\usepackage{amssymb}   
\usepackage{bm} 
\usepackage{dcolumn}
\usepackage{color}
\usepackage{mathrsfs}
\usepackage{amsfonts}
\usepackage{varioref}
\usepackage{textcomp}
\usepackage[normalem]{ulem}

\usepackage{multirow}
\usepackage{caption}
\usepackage{subcaption}
\RequirePackage[colorlinks,citecolor=blue,urlcolor=magenta,linkcolor=blue]{hyperref}
\allowdisplaybreaks
\labelformat{section}{Section #1} 
\labelformat{subsection}{Section #1} 
\labelformat{subsubsection}{Section #1}
\labelformat{subsubsubsection}{Section #1}
\labelformat{equation}{Eq.~(#1)} 
\labelformat{figure}{Fig.~#1} 
\labelformat{subfigure}{Fig.~\thefigure#1} 
\labelformat{table}{Table~#1} 
\labelformat{appendix}{Appendix #1}

\title{\bf Traversable Wormholes with a Spontaneous Symmetry Breaking Scalar Field}

\author{Soumya Chakrabarti\footnote{soumya.chakrabarti@vit.ac.in}$~^{1}$ and Chiranjeeb Singha\footnote{chiranjeeb.singha@iucaa.in}$~^{2}$\\
$^{1}$ {\small{School of Advanced Science, Vellore Institute of Technology
Vellore, Tiruvalam Rd, Katpadi, Tamil Nadu 632014
India}}\\
$^{2}${\small{Inter-University Centre for Astronomy and Astrophysics, Pune 411007, India}}\\
}

\date{\today}

\begin{document}
  
\maketitle

\begin{abstract} 

We argue that a spherically symmetric traversable wormhole solution of the Einstein field equations can be supported by minimally coupled self-interacting scalar field which allows a spontaneous symmetry breaking of the field around the wormhole throat. We study two cases : (i) the phantom wormhole solution of Bronnikov and (ii) a generalized Kiselev wormhole. We study the property of radial null geodesics and show that the metric can describe either a two-way or a one-way traversable wormhole depending on certain parameter ranges. The scalar field exhibits spontaneous symmetry breaking within the coordinate range where a wormhole throat forms and helps one suggest that spontaneous symmetry breaking may act as a threshold for wormhole throat formation. We also compute the radius of the photon sphere, the Lyapunov exponent, the shadow radius, and the innermost stable circular orbits for the geometries.

\end{abstract}

\section{Introduction}

Although the General Theory of Relativity (GR) is widely regarded as the most accurate description of gravity to date, the number of exact solutions to its field equations remains relatively small. This limitation is primarily due to the challenges involved in solving the field equations: coupled partial differential equations that form the basis of this theory. Even under spherical symmetry the field equations of GR are inherently quite restrictive. A clear example of this rigidity is found in the fact that even for a vacuum (where $T_{\mu\nu} = 0$), the Schwarzschild solution stands as the unique static and isotropic solution that satisfies the condition of asymptotic flatness \citep{1999physics...5030S, hilbert, weyl}. This result is central to the formulation of Jebsen-Birkhoff's theorem \citep{jebsen, birkhoff}. The Schwarzschild solution represents a singularity with zero proper volume surrounded by a horizon. When matter is incorporated into the field equations, they become even more restrictive and less predictable. Some partial solutions have been obtained in specific cases, such as GR coupled to self-interacting scalar fields of dilaton or axion types or when coupled to Maxwell electrodynamics or a Yang-Mills field. However, static field configurations in GR are heavily constrained outside the event horizon, as demonstrated by the no-hair theorems.  \\

A simple minimally coupled self-interacting scalar field $\phi$ contributes to the energy-momentum tensor through
\begin{equation}\label{minimallyscalar}
T^\phi_{\mu\nu} = \partial_\mu\phi\partial_\nu\phi - g_{\mu\nu}\Bigg[\frac{1}{2}g^{\alpha\beta}\partial_\alpha\phi\partial_\beta\phi - V(\phi)\Bigg]. 
\end{equation}
$V(\phi)$ is the self-interaction of the scalar field. The scalar field profile is imposed upon alongwith a requirement that in $r \rightarrow \infty$ limit the scalar field $\phi(r)$ should be a constant (one can choose to normalize this constant to zero). Moreover, a requirement of asymptotic flatness dictates that the self-interaction potential should have a local extremum at the origin of the scalar field. In fact, we can use these conditions to demonstrate the \textit{no-hair} theorem starting from the scalar field evolution equation

\begin{equation}
\partial_{\mu}\sqrt{-g}g^{\mu\nu}\partial_{\nu}\phi = \sqrt{-g}\frac{dV}{d\phi}.
\end{equation}

For a static case, only the $g^{rr}$ term contributes to the above equation. We multiply the equation with $\phi(r)$ and do integration by parts from the horizon to infinity, yielding

\begin{equation}
\left[\phi g^{rr} \sqrt{-g} \partial_{r}\phi \right]_{h}^{\infty} - \int_{h}^{\infty} dr \sqrt{-g} g^{rr} \left(\partial_{r}\phi \right)^{2} = \int_{h}^{\infty} dr \sqrt{-g} \phi \frac{dV}{d\phi}.
\end{equation}

Usually, one realizes a horizon $h$ at the largest zero of $g^{rr}$. Thus in principle it is possible to integrate the above equation for all $g^{rr} \geq 0$. If we assume regularity of $\phi$ and $\sqrt{-g}$ at the horizon, the boundary terms can be dropped (also keeping in mind an asymptotic fall-off of $\phi(r)$ as $r \rightarrow \infty$). In this context, the condition $\phi \frac{dV}{d\phi} \geq 0$ for all $r \geq h$ essentially implies that the self-interaction potential has a local minimum at $\phi = 0$ and no local maximum in the $\phi$-range considered outside the horizon. Under this condition, the integrands on both sides are non-negative. However, the integral on LHS brings in a negative sign, and for consistency, both sides of the equation should vanish identically. Thus the scalar field remains at $\phi(r) = \phi_0 = 0$, its asymptotic value and the resulting metric solution simply generates a Schwarzschild metric. This effectively rules out non-trivial scalar deformations of the Schwarzschild metric within a limit. The scalar \textit{no-hair} theorem demonstrated is now a popular avenue of research \citep{nh1, nh2, nh3, nh4, nh5}; however, there remains a curiosity if one can solve a system of static scalar field embedded within GR and weaken this constraint, even if slightly. \\

There are two further arguments which provide strong impetus behind the \textit{curious case of a scalar field} coupled to GR. The first argument is a motivation related to the study of cosmic expansion. In a spatially homogeneous cosmological universe, there is an omnipresent requirement of something beyond standard matter, capable of accommodating the idea of an accelerated expansion \citep{de1, de2, de3, de4, de5}. A scalar field with a suitably chosen interaction profile has served as the most popular candidate in this pursuit \citep{scalar1, scalar2}. The present scope of simple scalar field cosmological models appears to be bleak in light of the steep fifth force constraints on the baryon-quintessence interaction profile \citep{Adelberger:2003zx}; nevertheless, scalar fields remain capable of generating solutions of considerable interest. As a specific example, we refer to the pseudo-Nambu-Goldstone-Boson (pNGB) model \citep{PhysRevLett.75.2077} and models where a scalar field is capable of manipulating self-interaction via screening mechanisms \citep{PhysRevLett.93.171104, PhysRevLett.104.231301, 10.1093/mnras/stac1321}. \\

The second argument is the requirement of a (grand) unified theory of natural interaction, which should exhibit two crucial components: an underlying Lagrangian formalism to generate the equations of motion and the necessary fundamental couplings. In practice, these couplings do not have any direct derivations, and therefore, they are treated as constant parameters somehow related to the characteristic scale of interaction. The idea of variation of these couplings was proposed by Dirac as the \textit{`Large Numbers hypothesis’} \citep{Dirac}. There have been quite a few attempts to successfully accommodate this hypothesis within a theory of gravity using interacting scalar fields; however, all of them are partially complete. One of the well-established attempts is the field-theoretic description allowing variations in the Newtonian gravitational constant, the Brans-Dicke (BD) theory \citep{BD1, BD2, BD3}. In a BD theory, the scalar field is geometric (mass dimension two in natural units) and can be considered with or without any additional interaction. On the other hand, there are propositions of the theory of gravity accommodating the variation of $\alpha = \frac{e^{2}}{\hbar c}$, the fine structure constant. In such a theory, a scalar field interacts with the charged matter distribution of the universe and drives the variation of an $e$-field \citep{gamow, PhysRevD.25.1527}. Theories of this kind are usually associated with atleast a local violation of the equivalence principle, for example, a variation in electric charge $e$ leads to a breakdown of local charge conservation \citep{PhysRevD.25.1527}. There are quite a few attempts found in the literature to reproduce such variations within GR \citep{RevModPhys.75.403, 2011PThPh.126..993C} and overall, it is difficult to rule out the possibility of a common background mechanism that controls all such variation, although this has never been proved theoretically or experimentally. Intuitively, it can be a way forward to explore the features of a scalar field that can describe a variation in any of the fundamental couplings. Such a scalar field can not be termed \textit{exotic} as it has some motivations from fundamental interactions other than gravity. In this article, we work with a lagrangian where a minimally coupled self-interacting scalar field is capable of inducing spontaneous symmetry breaking. The self-interaction of the scalar field can be compared to a Higgs-like field with a small adjustment in the quadratic term,

\begin{equation}\label{potential}
V(\phi) = V_{0} + M(\phi) \phi^{2} + \lambda \phi^{4}.
\end{equation}

While one can consider $\lambda$ as a dimensionless parameter of the theory, the presence of $M(\phi)$ can lead to an evolving Higgs VEV $\nu$. It must be noted that $\phi$ need not be the standard model Higgs. We can imagine the standard model Higgs $\varphi$ having two components, the Higgs particle $h$ alongwith a classical background field $\phi$, with a corresponding Lagrangian given by

\begin{equation}
L =  - \frac{1}{2} \partial_{\mu} \varphi  \partial^{\mu} \varphi - \frac{\lambda}{4} (\varphi^{2} -\nu^{2})^{2} ~~,~~ \varphi = \nu_{0} + \phi + h = \nu(\phi) + h. 
\end{equation}

Due to this construction, the Higgs VEV $\nu$ can exhibit an evolution as a function of the classical field $\phi$. This variation plays an important role in the electroweak theory of interactions, where the Higgs boson is thought to be the mass generator of quarks. There is a linear correlation between the quark mass and Higgs VEV, $m_{e,q} = \lambda_{e,q} \nu$, where $\lambda_{e,q}$ is the Yukawa coupling \citep{GASSER198277, PhysRevLett.74.1071, Calmet:2002ja, FRITZSCH2009221}. While the quark masses are directly proportional, the VEV contribution to proton mass $m_{p}$ can be neglected since it depends primarily on the scale of quark-gluon interaction. As a consequence, any pre-assigned variation in VEV results in a variation of the proton-to-electron mass ratio $\mu$, which is directly measurable through the observation of molecular absorption spectra \citep{PhysRevLett.113.123002, PhysRevLett.113.210802, Sola:2016our, 10.1093/mnras/stab1910}. \\

In this article, we study the effect of introducing an $M(\phi)$ term in the scalar self-interaction $V(\phi) = V_{0} + M(\phi) \phi^{2} + \lambda \phi^{4}$ on two spherically symmetric static solutions of the Einstein field equations. The first metric tensor we discuss is a Phantom wormhole solution, initially derived by Bronnikov and Fabris for a minimally coupled scalar field with negative kinetic energy (sometimes favored cosmologically) \citep{Bronnikov_2006}. We further prove that this wormhole metric can be rewritten in a generalized Kiselev form. Kiselev's metric solution without any modification describes a black hole \cite{Kiselev_2003, Visser_2020}
\begin{equation}
ds^2 = - \left(1-{2m\over r} - {K\over r^{1+3w}} \right) dt^2 + {dr^2\over1-{2m\over r} - {K\over r^{1+3w}}} 
+ r^2 \,d\Omega_2^2 .
\end{equation}
This model can describe three classes of solutions within one framework: Schwarzschild for $w = 0$, Reissner-Nordstrom for $w = 1/3$, and Schwarzschild-(anti)-de~Sitter for $w = -1$. It has been proved that the Kiselev metric can not be supported by a perfect fluid or a simple quintessence. We regularize the metric tensor by replacing $r$ with $(r^2 + a^2)^{1/2}$ and make necessary adjustments in the other parameters to write it as
\begin{equation}
g_{00} = 1 - \frac{2 m}{\sqrt{a^2+r^2}} - \frac{p}{f(r) \left(a^2+r^2\right)^{\frac{1}{2} (3 w+1)}}.
\end{equation} 
We study the behavior of radial null geodesics for these two solutions and explain that depending on the regions of parameter space under consideration, the solutions behave as a traversable Wormhole. We show that it is possible to support these solutions using a scalar field which behaves as a constant in the asymptotic limit and exhibits spontaneous symmetry breaking (SSB) within the coordinate range associated with the formation of a wormhole throat. In other words, spontaneous symmetry breaking can be viewed as a threshold for the formation of a wormhole throat. Furthermore, we calculate the radius of the photon sphere, Lyapunov exponent, shadow radius, and innermost stable circular orbits for this spacetime and discuss their nature. The general requirement for non-trivial static solutions of the field equations of GR as well as of extensions of GR has always been steadfast, and there are quite a few solutions in literature (see for instance, \citep{Battista:2024gud, DiGrezia:2017daq, DeFalco:2020afv, DeFalco:2021klh, DeFalco:2021ksd, Chew:2024rin, Chew:2022enh, Chew:2023olq, Chew:2024bec, Chew:2024evh, Dzhunushaliev:2008bq, PhysRevD.109.044011, Pereira:2024gsl, Pereira:2024rtv, Bronnikov:2021liv}). Our construction can be applied to any of these solutions, and an appropriate association with SSB can be made. 

\section{The Phantom Wormhole: Solution and Properties}

The gravitational action we consider has the following form

\begin{equation}
S = \int d^{4}x \sqrt{-g} \left[ R + \frac{1}{2} g^{\mu \nu} \partial_{\mu} \phi \partial_{\nu} \phi + V(\phi) \right].
\end{equation}

The scalar field $\phi$ is a function of radial coordinate $r$ alone. The self-interaction of the scalar field is taken in a form as in \ref{potential}, i.e., $V(\phi) = V_{0} + \frac{1}{2} M(\phi) \phi^{2} + \frac{\lambda}{4} \phi^{4}$. We write a general static spherically symmetric metric tensor  

\begin{equation}\label{mainmetric}
ds^{2} = - G(r) dt^{2} + \frac{dr^{2}}{G(r)} + S(r)^{2} d\Omega^{2},
\end{equation}

Note, that $g_{11} = \frac{1}{g_{00}} = G(r)^{-1}$, therefore this metric is in a standard Schwarzschild form. The metric coefficients are functions of $r$ alone. The coordinates have natural domains, as in

\begin{equation}\label{mainmetricdomain}
r\in(-\infty,+\infty);\qquad 
t\in(-\infty,+\infty);\qquad
\theta\in [0,\pi];\qquad
\phi\in(-\pi,\pi].
\end{equation}

We proceed with an ansatz over the radius of two-sphere $S(r) = e^{-\sigma(r)}$. As we have already discussed in the introduction, $V_{0}$ will be treated as a parameter. $M(\phi)$ is expected to introduce a variation in the effective mass-like term for the scalar field $\phi$ and will be determined from the field equations of the theory. By varying the action with respect to the metric tensor and the scalar field $\phi$, we find the following set of coupled differential equations for the functions $\sigma(r)$, $G(r)$, and $\phi(r)$,

\begin{eqnarray}\label{feqgen}
&& \sigma'' - (\sigma')^{2} + \frac{1}{4} (\phi')^{2} = 0, \\&&
G'' - 2 G' \sigma' - V(\phi) = 0, \\&&
G'' - 2 G \left( 2 (\sigma')^{2} - \sigma'' \right) + 2 e^{2\sigma} = 0.
\end{eqnarray}

We write an exact solution to these field equations as

\begin{eqnarray}\label{exactG}
&& \sigma(r) = - \frac{1}{2} \ln(r^{2} + a^{2}), \\&&\label{exactG2}
G(r) = 1 + C_{1} + \frac{C_{2}}{2 a^{3}} \left[ a r + (a^{2} + r^{2}) \tan^{-1}(r/a) \right], \\&&
\phi(r) = C_{3} \pm 2 \tan^{-1}(r/a).
\end{eqnarray}

$a$ is a non-zero parameter of the solution. $C_{1}$, $C_{2}$ and $C_{3}$ are constants of integration. The functional form of $\sigma(r)$ exhibits logarithmic behavior at large distances, suggesting that the radial function falls off slowly. There can be two different profiles for the scalar field depending on the $\pm$ sign after the parameter $C_{3}$. Due to the presence of $\tan^{-1}(r/a)$ in the profile, the scalar field always asymptotes to constant values at large $r$. \\

\begin{figure}[t!]
\begin{center}
\includegraphics[angle=0, width=0.40\textwidth]{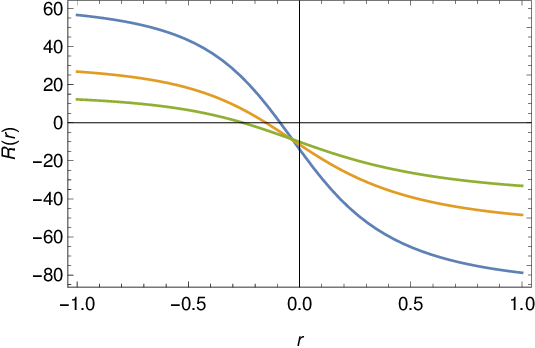}
\includegraphics[angle=0, width=0.40\textwidth]{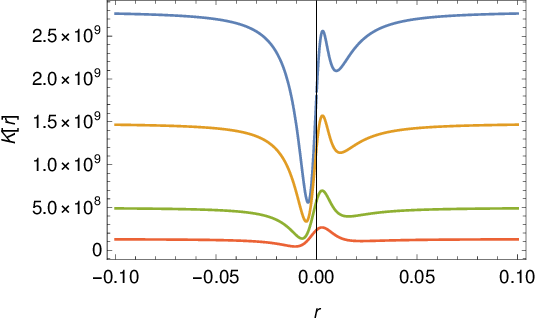}
\caption{Ricci scalar $R(r)$ and Kretschmann scalar $K(r)$ as a function of $r$ for different values of $a$. The parameter choices made in this graph are $C_{1} = 1$ and $C_{2} = 0.01$, while the parameter $a$ is varied.}
\label{fig_scalar}
\end{center}
\end{figure}

The first crucial property of this space-time metric is the regularity of curvature scalars. We calculate the Ricci and Kretschmann scalars as

\begin{eqnarray}
&& R = -\frac{6 a^7 C_{1}+2 a^5 \left(9 C_{1} r^2 + 1\right) + a^3 \left(12 C_{1} r^4 + 7 C_{2} r\right) + 3 C_{2} \left(a^4 + 3 a^2 r^2 + 2 r^4 \right) \tan^{-1}\left(\frac{r}{a}\right) + 6 a C_{2} r^3}{a^3 \left(a^2 + r^2 \right)^2}, \\&&\nonumber
K = \Big[3 C_{2}^2 \Big(a^2+r^2\Big)^2 \Big(a^4+2 a^2 r^2+2 r^4\Big) \tan^{-1}\Big(\frac{r}{a}\Big)^2 + 2 a C_{2} \Big(a^2+r^2\Big) \tan^{-1}\Big(\frac{r}{a}\Big) \Big(6 a^8 C_{1} + 2 a^6 \Big(9 C_{1} r^2 \\&&\nonumber
+ 2\Big) + a^4 r \Big(24 C_{1} r^3 + 5 C_{2} + 2 r\Big) + 2 a^2 \Big(6 C_{1} r^6 + 5 C_{2} r^3\Big) + 6 C_{2} r^5\Big) + a^2 \Big(12 a^{12} C_{1}^2 + 16 a^{10} C_{1} \Big(3 C_{1} r^2 + 1\Big) \\&&\nonumber
+ 4 a^8 \Big(21 C_{1}^2 r^4 + C_{1} r (5 C_{2} + 6 r) + 3\Big) + 4 a^6 \Big(C_{2} r \Big(15 C_{1} r^2 + 4\Big) + 2 C_{1} r^4 \Big(9 C_{1} r^2 + 1\Big) \Big) + a^4 r^2 \Big(24 C_{1}^2 r^6 \\&& 
+ 4 C_{2} \Big(16 C_{1} r^3 + r\Big) + 11 C_{2}^2\Big) + 2 a^2 C_{2} r^4 \Big(12 C_{1} r^3 + 7 C_{2}\Big) + 6 C_{2}^2 r^6\Big)\Big] \Big[a^6 \Big(a^2+r^2 \Big)^4 \Big]^{-1}.
\end{eqnarray} 

It is straightforward to note that the scalars do not diverge for any value of $r$. In particular, there is no singularity at $r = 0$ as long as $a$ is treated as a non-zero parameter and throughout this article we keep the parameter that way. The role of this parameter resembles the recently proposed solution of a regular black hole \citep{Simpson_2019} and its time-evolving analogue \citep{PhysRevD.104.024071}. We plot the scalars as a function of radial coordinate in \ref{fig_scalar} to prove our claim. For the purpose of this plot, we have chosen $C_{1} = 1$ and $C_{2} = 0.01$, while the parameter $a$ is being varied.  We also note that these scalars do not go to zero at an asymptotic limit $r \rightarrow \infty$, as it is found for a standard Schwarzschild solution. For our metric tensor both of the scalars reach a non-zero constant value at this limit, which is anyway not unexpected given the functional form of $g_{00} = 1 + C_{1} + \frac{C_{2}}{2 a^{3}} \left[ a r + (a^{2} + r^{2}) \tan^{-1}(r/a) \right]$. This is quite distinct in comparison to a Schwarzschild metric coefficient $\sim \left(1 - 2M/r \right)$, which can only go to $1$ in an asymptotic limit. For our solution, $g_{00}$ can reach a constant value (not equal to $1$) for all $a \neq 0$. In other words, the presence of the parameter $a$ encodes the departure from the standard Schwarzschild metric. \medskip

\begin{figure}[t!]
\begin{center}
\includegraphics[angle=0, width=0.40\textwidth]{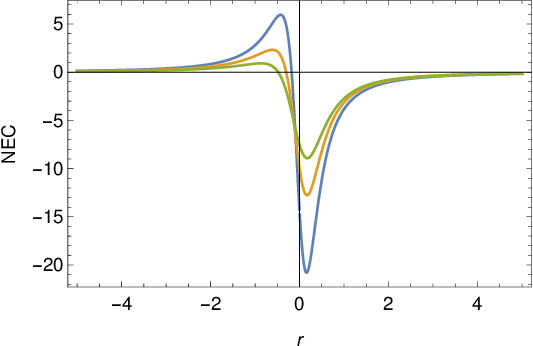}
\includegraphics[angle=0, width=0.40\textwidth]{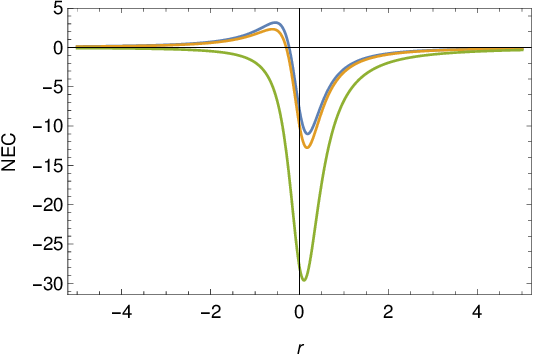}
\includegraphics[angle=0, width=0.40\textwidth]{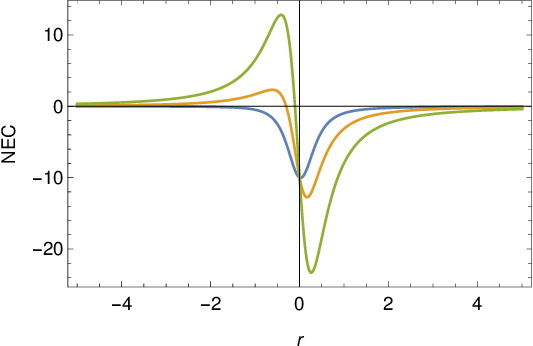}
\caption{Null Energy Condition as a function of $r$ for different values of $a$, $C_1$ and $C_2$. Top left : $C_{1} = C_{2} = 1$ while $a$ is being varied. Top right : $a = 0.5$ and $C_{2} = 1$, $C_1$ is being varied from $0.01$ to $1$. Below : $a = 0.5$ and $C_{1} = 1$ while $C_2$ is being varied from $0.1$ to $10$.}
\label{fig_NEC}
\end{center}
\end{figure}

Using the exact form of the metric coefficients, it is also straightforward to calculate the components of the energy-momentum tensor (using the mixed components of the field equations $G^{\mu}_{\nu} = 8 \pi G_N T^{\mu}_{\nu}$). The density is given by $\rho = - T^{t}_{t}$ and the radial pressure is given by $p_{r} = T^{r}_{r}$. Furthermore, we can check if there is any violation of Null Energy Condition (NEC) inside the spherical geometry. The necessary condition for an NEC to be valid is that the sum of density and radial pressure should be positive. Any violation of NEC is directly connected to the idea of a Null Convergence Condition (NCC), which is a characteristic of wormhole geometry, where radial null rays can defocus around the wormhole throat \citep{Visser:1995cc, PhysRev.48.73, MISNER1957525, Kolassis_1988, 10.10631.1666161}. We plot the NEC profile in \ref{fig_NEC} as a function of $r$. For the graph on the top left of \ref{fig_NEC}, we keep the two parameters $C_{1}$ and $C_{2}$ fixed at unity and vary the parameter $a$. For the graph on top right, we keep $a$ fixed at $0.5$ and $C_{2}$ fixed at $1$ while varying $C_1$ from $0.01$ to $1$. For the graph below, we keep $a$ fixed at $0.5$ and $C_{1}$ fixed at $1$ while varying $C_2$ from $0.1$ to $10$. \\

\begin{figure}[t!]
\begin{center}
\includegraphics[angle=0, width=0.40\textwidth]{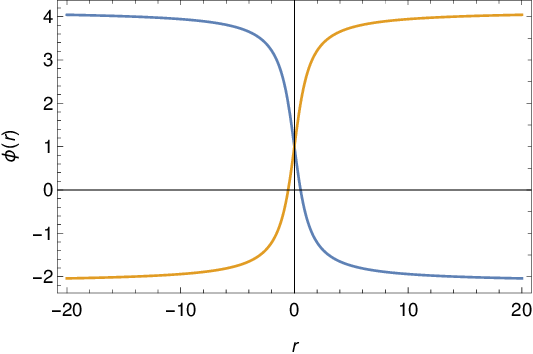}
\caption{$\phi(r)$ as a function of $r$. The blue curve shows the profile for a minus sign in $C_{3} \pm 2 \tan^{-1}(r/a)$, while the orange curve is for the plus sign. $C_3 = 1$.}
\label{fig_0}
\end{center}
\end{figure}

We plot the scalar field profile as a function of $r$ in \ref{fig_0}. The blue curve is the profile for a minus sign in $C_{3} \pm 2 \tan^{-1}(r/a)$, while the orange curve is for the plus sign. The constant $C_3$ is taken to be unity. It is seen that for both of the choices, the scalar field behaves as a constant at a large $r$. Finally, the solution for $M(\phi)$ is written straightaway from the field equations as

\begin{figure}[t!]
\begin{center}
\includegraphics[angle=0, width=0.40\textwidth]{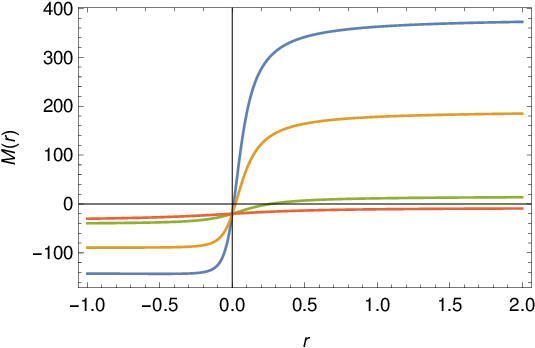}
\includegraphics[angle=0, width=0.40\textwidth]{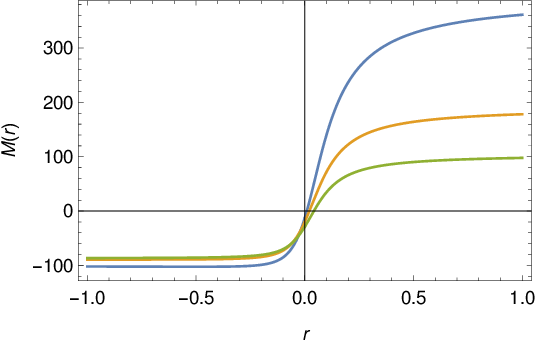}
\includegraphics[angle=0, width=0.40\textwidth]{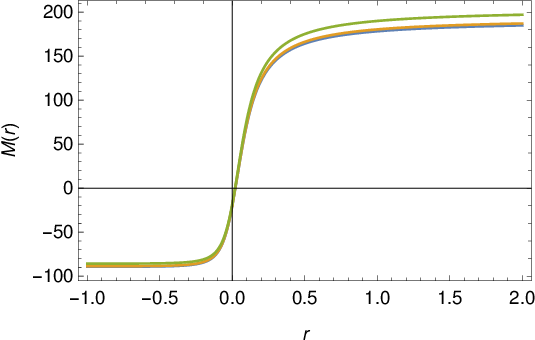}
\includegraphics[angle=0, width=0.40\textwidth]{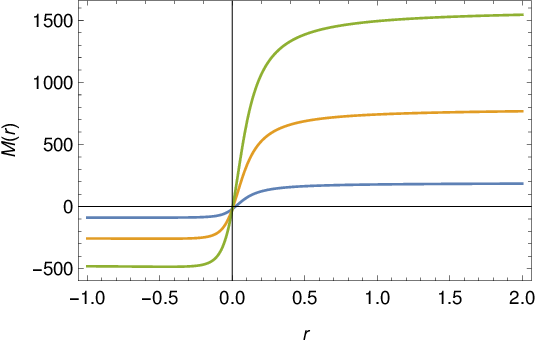}
\caption{Plot of $M(\phi)$ as a function of $r$ for $\phi(r) = C_{3} + 2 \tan^{-1}(r/a)$. Top left : different values of $a$ is considered while $C_{1}$, $C_{2}$ and $C_{3}$ are fixed. Top right : different values of $C_3$ is considered while $a$, $C_{1}$ and $C_{2}$ are fixed. Bottom left : different values of $C_1$ is considered while $a$, $C_{2}$ and $C_{3}$ are fixed. Bottom right : different values of $C_2$ is considered while $a$, $C_{1}$ and $C_{3}$ are fixed.}
\label{fig_2}
\end{center}
\end{figure}
  
\begin{equation}
M = 2 \bigg\{G'' +\frac{2 G' r}{(r^{2}+a^{2})} - V_{0}-\frac{\lambda}{4}\phi^{4}\bigg\}~.
\end{equation}

\begin{figure}[t!]
\begin{center}
\includegraphics[angle=0, width=0.40\textwidth]{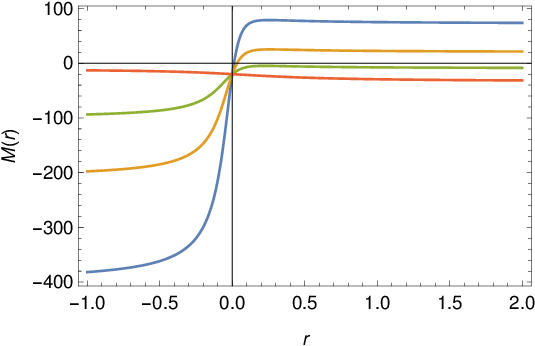}
\includegraphics[angle=0, width=0.40\textwidth]{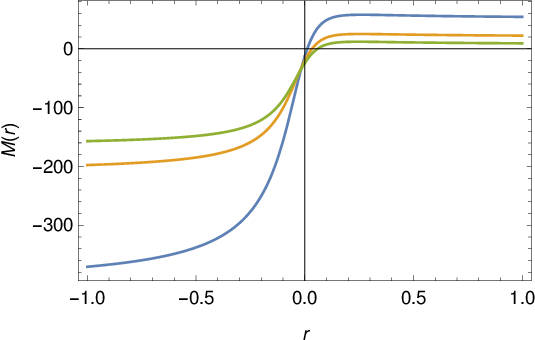}
\includegraphics[angle=0, width=0.40\textwidth]{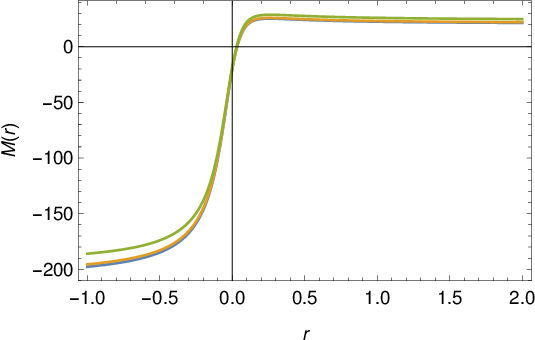}
\includegraphics[angle=0, width=0.40\textwidth]{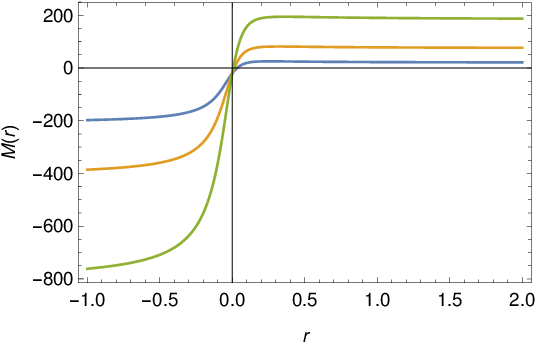}
\caption{Plot of $M(\phi)$ as a function of $r$ for $\phi(r) = C_{3} - 2 \tan^{-1}(r/a)$. Top left : different values of $a$ is considered while $C_{1}$, $C_{2}$ and $C_{3}$ are fixed. Top right : different values of $C_3$ is considered while $a$, $C_{1}$ and $C_{2}$ are fixed. Bottom left : different values of $C_1$ is considered while $a$, $C_{2}$ and $C_{3}$ are fixed. Bottom right : different values of $C_2$ is considered while $a$, $C_{1}$ and $C_{3}$ are fixed.}
\label{fig_4}
\end{center}
\end{figure}

It is important to note that the solution we have found is consistent if and only if $M(\phi)$ is a function. If it is a constant, then the requirement that $[G'' + \frac{2 G' r}{(r^{2}+a^{2})} - V_{0} - \frac{\lambda}{4}\phi^{4}]$ introduces a steep constraint on the allowed solution spectrum. In \ref{fig_2} we plot $M(\phi)$ as a function of $r$ for $\phi(r) = C_{3} + 2 \tan^{-1}(r/a)$. The graph on the top left is for different values of $a$ while the three other parameters $C_{1}$, $C_{2}$ and $C_{3}$ are fixed. The graph on top right is for different values of $C_3$ while $a$, $C_{1}$ and $C_{2}$ are fixed. The graph on bottom left is for different values of $C_1$ while $a$, $C_{2}$ and $C_{3}$ are fixed. The graph on bottom right is for different values of $C_2$ while $a$, $C_{1}$ and $C_{3}$ are fixed. Similarly, in \ref{fig_4} we plot $M(\phi)$ as a function of $r$ for the second scalar field profile, $\phi(r) = C_{3} - 2 \tan^{-1}(r/a)$. The graph on the top left is for different values of $a$ while the three other parameters $C_{1}$, $C_{2}$ and $C_{3}$ are fixed. The graph on top right is for different values of $C_3$ while $a$, $C_{1}$ and $C_{2}$ are fixed. The graph on bottom left is for different values of $C_1$ while $a$, $C_{2}$ and $C_{3}$ are fixed. The graph on bottom right is for different values of $C_2$ while $a$, $C_{1}$ and $C_{3}$ are fixed. Irrespective of the choice of $\phi$ profile, the qualitative behavior of $M(\phi)$ remains the same as a function of $r$. There is always a switch from negative into positive values of $M(\phi)$ within a small neighbourhood of $r = 0$, but never exactly at $r = 0$. In a sense, the solution therefore behaves as in a \textit{symmetron} where the interaction potential is taken as
\begin{equation} 
V(\phi) = -{1\over 2}\mu^2\phi^2+{1\over 4}\lambda\phi^4 ~~,~~ A(\phi)=1+{1\over 2M^2}\phi^2\,,
\label{ourpotential} 
\end{equation}

where $A(\phi)$ defines a universal coupling of the scalar field to the space-time metric in an Einstein frame \citep{PhysRevLett.104.231301, PhysRevD.82.083503, PhysRevD.84.103521}. The effective potential in this case becomes dependent on $\mu$, $M$ (two mass scales) and the constant $\lambda$, written as
\begin{equation} 
V_{\rm eff}(\phi)={1\over 2}\left({\rho\over M^2}-\mu^2\right)\phi^2+{1\over 4}\lambda\phi^4\,.
\end{equation}

Due to the negative mass term of $V(\phi)$ there is a spontaneous breakdown of ${\bf Z_2}$ symmetry. The sign of the quadratic term in the effective potential depends on the local matter density. As a result, the spontaneous breaking of ${\bf Z_2}$ symmetry is also dependent on the presence of the source. When $\rho = 0$, which corresponds to the region outside the source, the potential spontaneously breaks reflection symmetry, and the scalar field acquires a vacuum expectation value (VEV) $\phi_0 \equiv \mu/\sqrt{\lambda}$. However, within the source, if the parameters are chosen such that $\rho > M^2 \mu^2$, the effective potential no longer induces symmetry breaking, and the VEV becomes zero. The present solution, however, also allows a few cases, where there is no transition from negative into positive, i.e., the spontaneous symmetry always remains broken throughout the coordinate range $r\in(-\infty,+\infty)$. \\

We now look into the radial null curves by setting the spacetime interval to zero. On a fixed plane ($d\theta = d\phi = 0$) such a curve can describe the path of a massless particle, such as a photon, moving radially through the spacetime, obeying

\begin{equation}\label{nullgeo}
\frac{dr}{dt} = \pm \left[ 1 + C_{1} + \frac{C_{2}}{2 a^{3}} \left( a r + (a^{2} + r^{2}) \tan^{-1}(r/a) \right) \right].
\end{equation}

One can define a coordinate speed of light in the spacetime metric as $c(r) = \left\vert\frac{dr}{dt}\right\vert$. The coordinate speed of light $c(r)$ depends on the radial coordinate $r$ and is naturally affected by the curvature of spacetime. In flat, uncurved spacetime, the speed of light is supposed to be constant, but in curved spacetime, the speed of light should vary depending on the location within the gravitational field. An effective refractive index $n(r)$ can also be defined as the inverse of the coordinate speed of light, $n(r) = \frac{1}{c(r)}$. It describes how the curvature of spacetime affects the propagation of light. If $c(r)$ is reduced compared to the standard value of the speed of light, the refractive index $n(r)$ is increased, indicating that light effectively slows down as it passes through these regions. Effectively, the function $n(r)$ provides a measure of the gravitational lensing effects in the spacetime. Gravitational lensing occurs because light rays are bent and slowed down as they pass through regions of spacetime with strong curvature, which is analogous to the way light bends and slows down when passing through a medium with a varying refractive index in optics. \\

\begin{figure}[t!]
\begin{center}
\includegraphics[angle=0, width=0.40\textwidth]{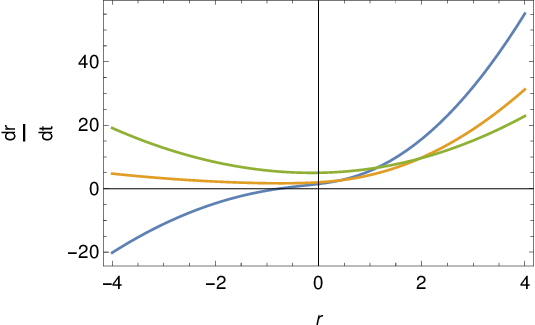}
\includegraphics[angle=0, width=0.40\textwidth]{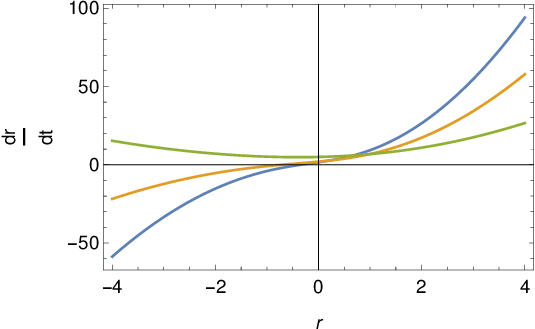}
\includegraphics[angle=0, width=0.40\textwidth]{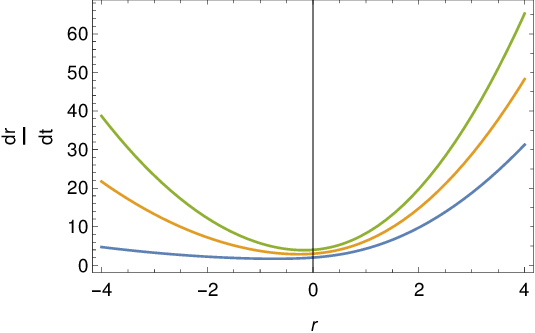}
\includegraphics[angle=0, width=0.40\textwidth]{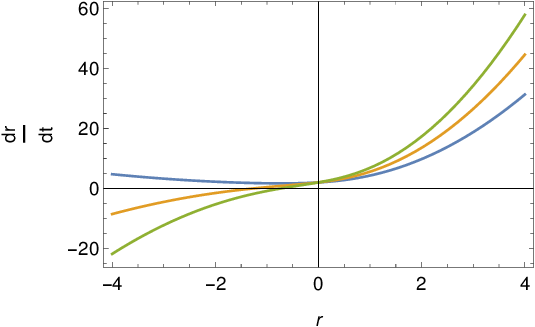}
\caption{$\frac{dr}{dt}$ as a function of $r$ for different set of initial conditions. Top left : $C_1 = 1, C_2 = 1$ and the value of $a$ is varied. Top right : $C_1 = 1, C_2 = 3$ and the value of $a$ is varied. Bottom left : $a = C_2 = 1$, while the parameter $C_1$ is varied. Bottom right : $a = C_1 = 1$, while the parameter $C_2$ is varied.}
\label{fig_1}
\end{center}
\end{figure}

We use \ref{nullgeo} to check if there is any formation of the horizon for any value or range of $r$, for which we simply require that $\frac{dr}{dt} = 0$ for any $r \in (-\infty, +\infty)$. Given the fact that \ref{nullgeo} is not straightforward to solve, primarily due to the presence of a $\tan^{-1}$ term, we plot the RHS of the equation in \ref{fig_1}. The graph on the top left is for a chosen set of $C_1 = 1$, $C_2 = 1$. We change the value of $a$ and find that as long as $a \lesssim 0.9$, the curve intersects zero, and we can, in principle, have a horizon at some coordinate location. However, this is definitely not a black hole since we do not have any curvature singularities. It can be rendered as a one-way wormhole with an extremal null throat. However, for $a \gtrsim 0.9$, we have $\frac{dr}{dt}\neq 0$ for any $r \in (-\infty, +\infty)$ and therefore the metric represents a two-way traversable wormhole. For this set, $a \sim 0.9$ defines a threshold of horizon formation, however, this is not an exhaustive analysis. For a different set $C_1 = 1$, $C_2 = 3$ (graph on top right), the threshold value of $a$ is found to be $\sim 1.25$. The bottom left graph is for a fixed set of $a = C_2 = 1$, while the parameter $C_1$ is varied, and for this set, we do not have any formation of horizon. The bottom right graph is for a fixed set of $a = C_1 = 1$, while the parameter $C_2$ is varied. \\

In order to have a mathematical argument behind this parameter-dependence of horizon formation, we approximate \ref{nullgeo} around a very small neighborhood of $r = 0$ (essentially to ignore any terms $\sim \textit{O}(r^3)$) or higher. Under this approximation $\tan^{-1}(\frac{r}{a}) \approx \frac{r}{a}$, and the radial null geodesic curve equation becomes a quadratic equation
\begin{equation}
C_1 {r_h}^2 + \frac{C_2}{a^2}{r_h} + (1 + C_1 a^2) = 0.
\end{equation}  
Since we are expecting the coordinate location of a \textit{horizon}, it is better to use ${r_h}$ in place of $r$. By the standard method of treating quadratic equations, this must have two roots defined as
\begin{equation}
{r_h} = - \frac{C_2}{a^2} \pm \frac{\sqrt{{C_2}^2 - 4a^4 C_1 (1 + C_1 a^2)}}{2a^2 C_1}.
\end{equation}

Therefore we make a deduction that the realistic coordinate location of the horizon can only be found as long as ${C_2}^2 - 4a^4 C_1 (1 + C_1 a^2) \geq 0$ or ${C_2}^2 \geq 4a^4 C_1 (1 + C_1 a^2)$. \\

It can be proved that the parameter $a$ is nothing but the coordinate value at which a wormhole throat forms. This is evident through a straightforward coordinate transformation, $r^{2} + a^{2} = l^2$, applied to the metric in \ref{mainmetric}. The transformation leads to a domain for the radial coordinate $l\in(a, +\infty)$, indicating that now $a$ can be treated as the minimum value of the radial coordinate. As a consequence, the metric tensor in terms of $l$ takes on a new form

\begin{equation}\label{newmetric}
ds^{2} = - G_{1}(l) dt^{2} + \frac{dr^{2}}{\left(1 - \frac{a^2}{l^2}\right)G_{1}(l)} + l^{2} d\Omega^{2},
\end{equation}

In this framework, the modified $g_{11}$ term resembles a wormhole throat function \citep{PhysRevD.104.024071}. A wormhole throat is defined as a closed, space-like two-dimensional surface with minimal area. To analyze this, one has to examine an embedding geometry of the metric on a constant time slice, where $\theta = \pi/2$. This leads to
\begin{equation}\label{3_d}
d\sigma^2 = B^{2} dl^{2} + l^{2} d\phi^{2} ~~;~~ B^{2} = \frac{1}{\left(1 - \frac{a^2}{l^2}\right)G_{1}(l)}.
\end{equation}

$d\sigma^2$ can be mapped to a generic cylindrically symmetric metric in $3$-dimension
\begin{equation}\label{R3d}
d\sigma^2 = dz^2 + d\rho^2 + \rho^2 d\phi^2,
\end{equation}

defined on the surface of revolution $\rho(z)$. Comparing these two equations we establish the correlation
\begin{equation}
\rho^2 = l^{2} ~;~ dz^{2} + d\rho^{2} = B^{2}dl^{2},
\end{equation}

leading further to the conditions 
\begin{equation}\label{ee2}
\frac{d\rho}{dz} = \frac{1}{(B^{2}-1)^{\frac{1}{2}}} ~~;~~ \frac{d^2\rho}{dz^2} = -\frac{\frac{dB}{dl}B}{(B^{2}-1)^{2}}.
\end{equation}

In a cylindrically symmetric coordinate system, the throat of a wormhole can be visualized in an embedding diagram as a circle of radius $\rho$ on a surface of revolution. At the throat, the radius $\rho(z)$ should reach its minimum. A straightforward calculation can be done to show that in the limit $l \rightarrow a$, $\rho(z)$ indeed attains its minimum, satisfying the following two conditions

\begin{equation}\label{ee4}
\frac{d\rho}{dz}\Big|_{l \rightarrow a} = 0 ~~;~~ \frac{d^2\rho}{dz^2}\Big|_{l \rightarrow a} > 0.
\end{equation}

\section{The Generalized Kiselev Wormhole: Solution and Properties}
Using a simple algebraic manipulation one can rewrite the Phantom wormhole metric in \ref{exactG2} as
\begin{equation}\label{exactGK}
G(r) = 1-\frac{2 m}{\sqrt{a^2+r^2}}-\frac{p(r)}{\left(a^2+r^2\right)^{\frac{1}{2} (3 w+1)}},
\end{equation}
where
\begin{equation}
p(r) = C_1 \left(a^2+r^2\right)^{\frac{3 (w+1)}{2}}-2 m \left(a^2+r^2\right)^{\frac{3 w}{2}}+\frac{C_2 \left(a^2+r^2\right)^{\frac{3 w}{2}+\frac{1}{2}} \left\lbrace a^2 \tan ^{-1}\left(\frac{r}{a}\right)+r^2 \tan ^{-1}\left(\frac{r}{a}\right)+a r \right\rbrace}{2 a^3}.
\end{equation}
This is nothing but a generalization of the Kiselev metric \citep{Kiselev_2003} for which the function $p(r)$ is a number. Note that the parameter $w$ remains unperturbed, which, in the standard Kiselev metric, allows one to toggle between three classes of solutions, namely, Schwarzschild for $w = 0$, Reissner-Nordstrom for $w = 1/3$ and Schwarzschild-(anti)-de~Sitter for $w=-1$. Since there seems to be a way to associate an SSB-breaking scalar field with the Phantom wormhole, it is a natural curiosity to ask if the same can also be done for a Kiselev metric or its generalization. We propose the following generalization
\begin{equation}\label{exactGGK}
G(r) = 1-\frac{2 m}{\sqrt{a^2+r^2}} - \frac{p}{f(r) \left(a^2+r^2\right)^{\frac{n(r)}{2} }},
\end{equation} 
where
\begin{eqnarray}\nonumber
&& \sigma(r) = - \frac{1}{2} \ln(r^{2} + a^{2}) ~~,~~ \phi(r) = C_{3} \pm 2 \tan^{-1}(r/a) ~~,~~ n(r) = (3w+1), \\&&\nonumber
f(r) = a^{-2} p^{-1}-a^2 C_1 p r^2 \left(a^2+r^2\right)^{\frac{3 w}{2}+\frac{1}{2}}+a^2 C_2 p r^2 \left(a^2+r^2\right)^{\frac{3 w}{2}+\frac{1}{2}} \tan ^{-1}\left(\frac{r}{a}\right)-2 a^2 m \left(a^2+r^2\right)^{\frac{3 w}{2}}- \\&&\nonumber 
m r^2 \sqrt{\frac{a^2+r^2}{a^2}} \left(a^2+r^2\right)^{\frac{3 w}{2}}-a^2 m \sqrt{\frac{a^2+r^2}{a^2}} \left(a^2+r^2\right)^{\frac{3 w}{2}}-a^4 C_1 p \left(a^2+r^2\right)^{\frac{3 w}{2}+\frac{1}{2}} \\&&\nonumber
+a^4 C_2 p \left(a^2+r^2\right)^{\frac{3 w}{2}+\frac{1}{2}} \tan ^{-1}\left(\frac{r}{a}\right)+a^3 C_2 p r \left(a^2+r^2\right)^{\frac{3 w}{2}+\frac{1}{2}}.
\end{eqnarray}

\begin{figure}[t!]
\begin{center}
\includegraphics[angle=0, width=0.40\textwidth]{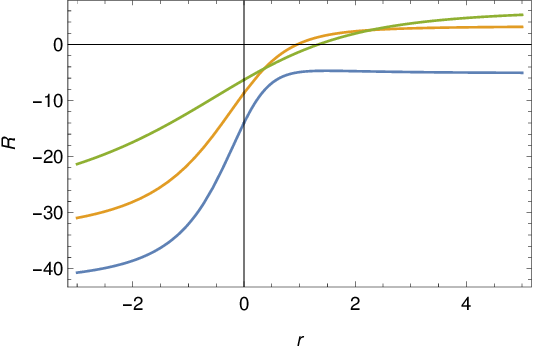}
\includegraphics[angle=0, width=0.40\textwidth]{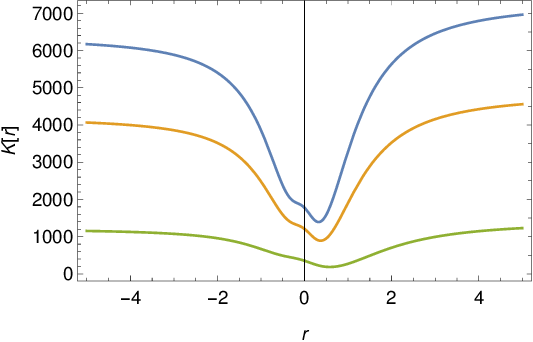}
\caption{Ricci scalar $R(r)$ and Kretschmann scalar $K(r)$ as a function of $r$ for different values of $a$ for a generalized Kiselev metric}
\label{fig_scalar_k}
\end{center}
\end{figure}

The above set of expressions can be found by solving the field equations defined in \ref{feqgen} alongwith the metric ansatz \ref{exactGGK}. The first thing we check and confirm is that the curvature scalars, namely the Ricci and Kretschmann scalar do not diverge for any value of $r$. In particular, there is no singularity at $r = 0$ as long as $a$ is treated as a non-zero parameter. Once again this parameter can be treated as a wormhole throat radius. We plot the scalars as a function of radial coordinate in \ref{fig_scalar_k}, choosing particular values of $C_{1}$, $C_{2}$, $m$ and $p$, varying the parameter $a$ alone. The qualitative behavior of the scalars remains similar even for different values of the parameters. We note that these scalars become constant at an asymptotic limit $r \rightarrow \infty$, however, this constant can be scaled to zero for an appropriately chosen set of parameter values.
  
\begin{eqnarray}\nonumber
&& R \Rightarrow  \Big(-\frac{2 m}{a^3}-2 C_1 p + \pi C_2 p \Big), \\&&\nonumber
K \Rightarrow \frac{60 p (2 C_1 -\pi  C_{2})}{a^3} + \frac{4}{a^6} \Bigg\lbrace\frac{98}{\Big(a^3 p (2 C_1 -\pi  C_{2}) + 2\Big)^2} - \frac{2}{a^3 p (2 C_1 - \pi C_{2}) + 4} + \frac{1}{\Big(a^3 p (2 C_1 -\pi C_{2}) + 4\Big)^2} \\&&\nonumber
- \frac{16}{a^3 p (2 C_1 - \pi C_{2})+6} + \frac{16}{\Big(a^3 p (2 C_1 -\pi C_{2}) + 6\Big)^2}-\frac{54}{a^3 p (2 C_1 - \pi C2)+8}+\frac{81}{\Big(a^3 p (2 C1-\pi C2) + 8\Big)^2} \\&&\nonumber
+\frac{72}{a^3 p (2 C_1 -\pi C_{2})+2}+73 \Big\rbrace+6 p^2 (\pi C_2 -2 C_{1})^2 . 
\end{eqnarray}

\begin{figure}[t!]
\begin{center}
\includegraphics[angle=0, width=0.40\textwidth]{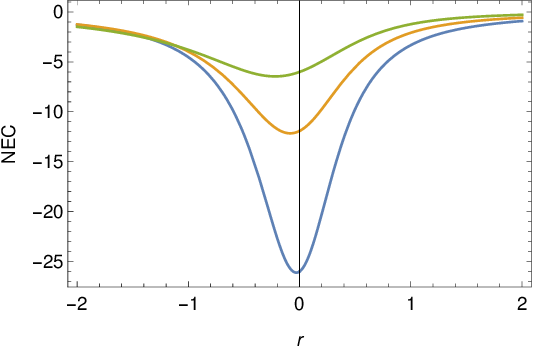}
\includegraphics[angle=0, width=0.40\textwidth]{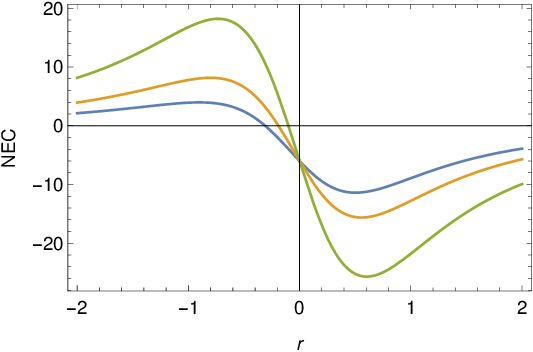}
\caption{Null Energy Condition as a function of $r$ for different values of $C_2$. $C_{1} = C_{2} = 1$, $a = 0.5$. Evolution for positive and negative values of $C_{2}$ are shown on left and right.}
\label{fig_NEC_k}
\end{center}
\end{figure}

One can calculate the components of the energy-momentum tensor for this metric using the field equations $G^{\mu}_{\nu} = 8 \pi G_N T^{\mu}_{\nu}$. We study, in particular, the NEC defined as $- T^{t}_{t} + T^{r}_{r}$ and plot the profile in \ref{fig_NEC_k} as a function of $r$. The graph on the left is for positive values of $C_{2}$, and the one on the right is for negative $C_{2}$. It is clearly seen that the NEC is violated at least in the present universe, i.e., in the coordinate range $0 < r < \infty$. \\

\begin{figure}[t!]
\begin{center}
\includegraphics[angle=0, width=0.40\textwidth]{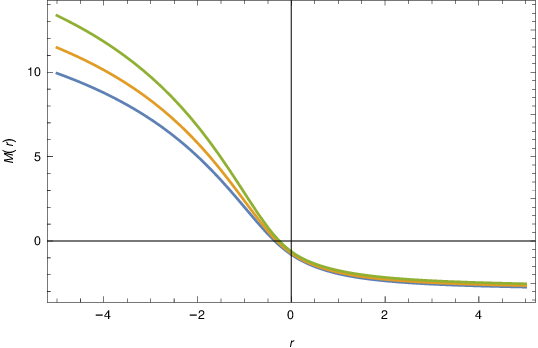}
\includegraphics[angle=0, width=0.40\textwidth]{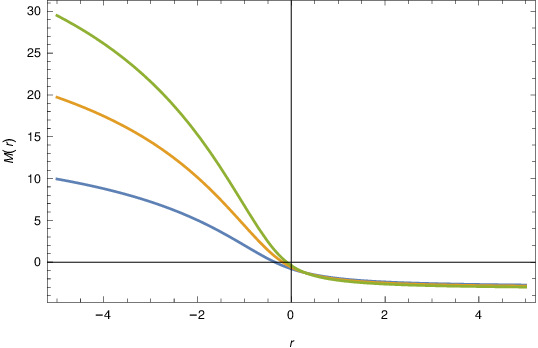}
\includegraphics[angle=0, width=0.40\textwidth]{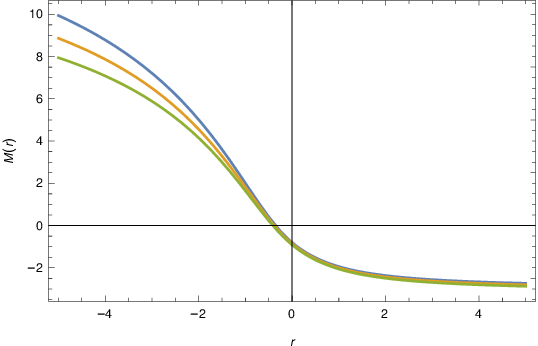}
\includegraphics[angle=0, width=0.40\textwidth]{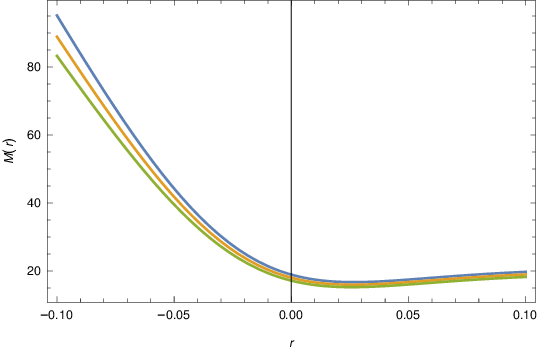}
\caption{Plot of $M(\phi)$ as a function of $r$ for $\phi(r) = C_{3} + 2 \tan^{-1}(r/a)$. Top left: different values of $m$ are considered while the other parameters are fixed. Top right: different values of $p$ are considered while the other parameters are fixed. Bottom left: different values of $C_3$ are considered while the other parameters are fixed. Bottom right : evolution for $C_3 >> 0$.}
\label{fig_mass_k}
\end{center}
\end{figure}

We note once again that the solutions discussed in the article are consistent if and only if the term $M(\phi)$ in the self-interaction potential is a function. Using \ref{feqgen} and \ref{exactGGK}, we solve for $M(\phi)$, and in \ref{fig_mass_k}, we plot it as a function of $r$. The graph on the top left is for different values of $a$ while the three other parameters $C_{1}$, $C_{2}$ and $C_{3}$ are fixed. The graph on the top left shows $M(r)$ for different values of $m$ while the other parameters are fixed. The graph on top right shows $M(r)$ for different values of $p$ while the other parameters are fixed. The bottom left graph shows $M(r)$ profile for different values of $C_3$. There is always a switch from negative into positive values of $M(\phi)$ within a small neighborhood of $r = 0$, but never exactly at $r = 0$, signifying the spontaneous breakdown of $Z_2$ symmetry. However, unlike a Phantom wormhole metric, it is possible in this case to have a solution with no SSB. This can be seen by taking values of $C_{3} >> C2$. The $M(r)$ profile corresponding to this case is shown in the bottom right graph of \ref{fig_mass_k}. \\

\begin{figure}[t!]
\begin{center}
\includegraphics[angle=0, width=0.40\textwidth]{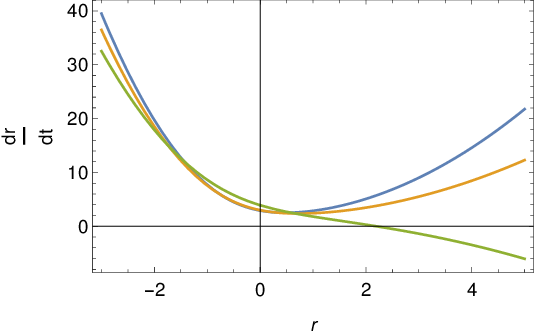}
\includegraphics[angle=0, width=0.40\textwidth]{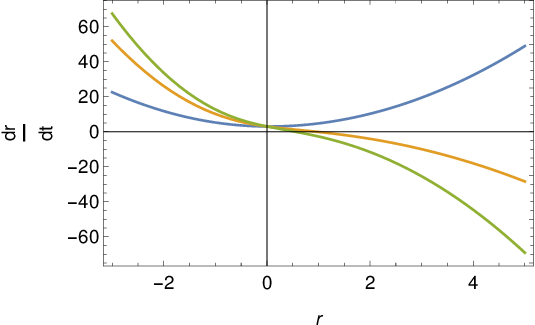}
\includegraphics[angle=0, width=0.40\textwidth]{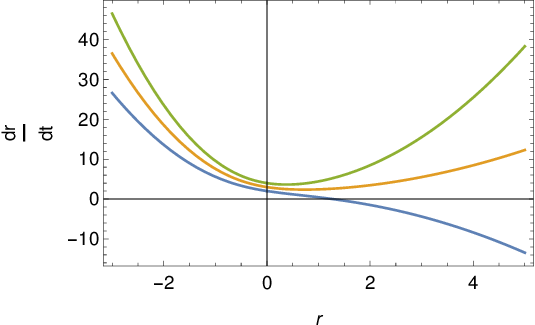}
\includegraphics[angle=0, width=0.40\textwidth]{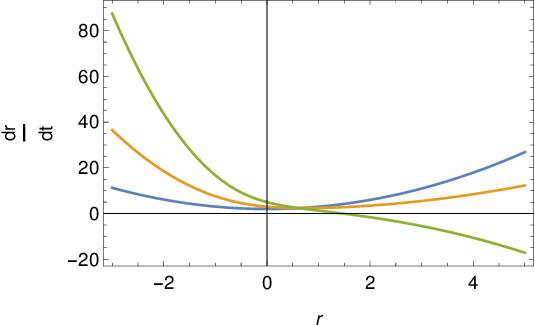}
\caption{$\frac{dr}{dt}$ as a function of $r$ for different set of initial conditions. Top left: The value of $a$ is varied while the other parameters are fixed. Top right: The value of $C_{2}$ is varied while the other parameters are fixed. Bottom left: The parameter $m$ is varied while the other parameters are fixed. Bottom right: The parameter $p$ is varied while the other parameters are fixed.}
\label{null_k}
\end{center}
\end{figure}

We study the evolution of $g_{00}$ component of the generalized Kiselev metric to see if $\frac{dr}{dt} = 0$ for any value of $r \in (-\infty, +\infty)$, which marks the formation of a horizon. We plot $\frac{dr}{dt}$ as a function of $r$ in \ref{null_k}. It is straightforward to deduce that there are no zeros of $\ \frac {dr}{dt}$ for most of the parameter space. The metric tensor describes a two-way traversable wormhole apart from a few exceptions where one might have a formation of the horizon without any central singularity at $r = 0$ or a regular black hole. 

\section{Photon sphere, Lyapunov exponent, Shadow radius, and Innermost Stable Circular Orbits}

In this section, we discuss a few additional properties of the two classes of space-time metrics described here: the radius of the photon sphere $(r_{\rm ph})$, the shadow radius ($r_{sh}$), the innermost stable circular orbits $(r_{\textrm{ISCO}})$ and the Lyapunov exponent $(\lambda)$. While a photon sphere is associated with massless particles, an ISCO is related to massive particles \citep{Virbhadra:1999nm, PhysRevD.65.103004, PhysRevD.77.124014, 10.1063/1.1308507, PhysRevD.79.083004}. In particular, the existence of a photon circular orbit plays a crucial role in understanding the characteristic signatures of the quasi-normal modes and shadow radius. We parametrize the tangent vector to the worldline of any particle (massless or massive) by an affine parameter $\tau$ and write
\begin{equation}
g_{ab}\frac{dx^{a}}{d\tau}\frac{dx^{b}}{d\tau}=-g_{tt}\left(\frac{dt}{d\tau}\right)^{2}+g_{rr}\left(\frac{dr}{d\tau}\right)^{2}+\left(r^{2}+a^{2}\right)\left\lbrace\left(\frac{d\theta}{d\tau}\right)^{2}+\sin^{2}\theta \left(\frac{d\phi}{d\tau}\right)^{2}\right\rbrace \ .
\end{equation}

The cases we are interested in are timelike (massive particle) and null (massless particle) and are classified as  
\begin{equation}
ds^{2}/d\lambda^2=\epsilon ~~,~~    
\epsilon = \left\{
    \begin{array}{rl}
    -1 & \qquad\mbox{timelike worldline} \\
     0 & \qquad\mbox{null worldline} .
    \end{array}\right. 
\end{equation}

For a spherically symmetric metric once again we take $\theta = \frac{\pi}{2}$ and write equations on an equatorial plane as

\begin{equation}\label{1}
    g_{ab}\frac{dx^{a}}{d\lambda}\frac{dx^{b}}{d\lambda}=-g_{tt}\left(\frac{dt}{d\lambda}\right)^{2}+g_{rr}\left(\frac{dr}{d\lambda}\right)^{2}+\left(r^{2}+a^{2}\right)\left(\frac{d\phi}{d\lambda}\right)^{2}=\epsilon \ .
\end{equation}

Using the Killing symmetries we can express $\frac{dt}{d\lambda}$ and $\frac{d\phi}{d\lambda}$ in terms of conserved energy $E$ and angular momentum $L$ per unit mass \citep{Wald:1984rg} as
\begin{equation}\label{2}
    G(r)\left(\frac{dt}{d\lambda}\right)=E \ ; \qquad\quad \left(r^{2}+a^{2}\right)\left(\frac{d\phi}{d\lambda}\right)=L.
\end{equation}

So far the mathematical construction does not make any particular choice of $G(r)$. We first discuss the generic calculations and remind the reader that the functional forms of $G(r)$ for the two solutions are given in \ref{exactG} (for a Phantom wormhole) and \ref{exactGGK} (for a Kiselev wormhole). Combining \ref{1} and \ref{2} we derive

\begin{equation}
\left(\frac{dr}{d\lambda}\right)^{2} = E^{2} + G(r)\left\lbrace\epsilon-\frac{L^{2}}{r^{2}+a^{2}}\right\rbrace,
\end{equation}

from which we can write an effective potential for any geodesic orbit as follows

\begin{equation}\label{effpot}
V_{\epsilon}(r) = G(r)\left\lbrace -\epsilon+\frac{L^{2}}{r^{2}+a^{2}}\right\rbrace \ .
\end{equation}

For a massless particle $\epsilon = 0$. We look for the $r$ values which can generate $V_{0}^{'}(r) = 0$. Under spherical symmetry, these points define the coordinate locations sufficiently close to the mass enclosed within the metric, where photons are restricted to travel along circular geodesic orbits. The effective potential for such a case is written as $V_{0}(r) = G(r)\left(\frac{\mathbb{L}^2}{r^2+a^{2}}\right)$. The solution $r_{\rm ph}$ of the equation $V_{0}'(r) = 0$ gives us the radius of the circular photon orbit \citep{Chakraborty:2021dmu, Mishra:2019trb, Berry:2020ntz, Tang:2017enb}. \\ 
\begin{figure}[t!]
\begin{center}
\includegraphics[angle=0, width=0.40\textwidth]{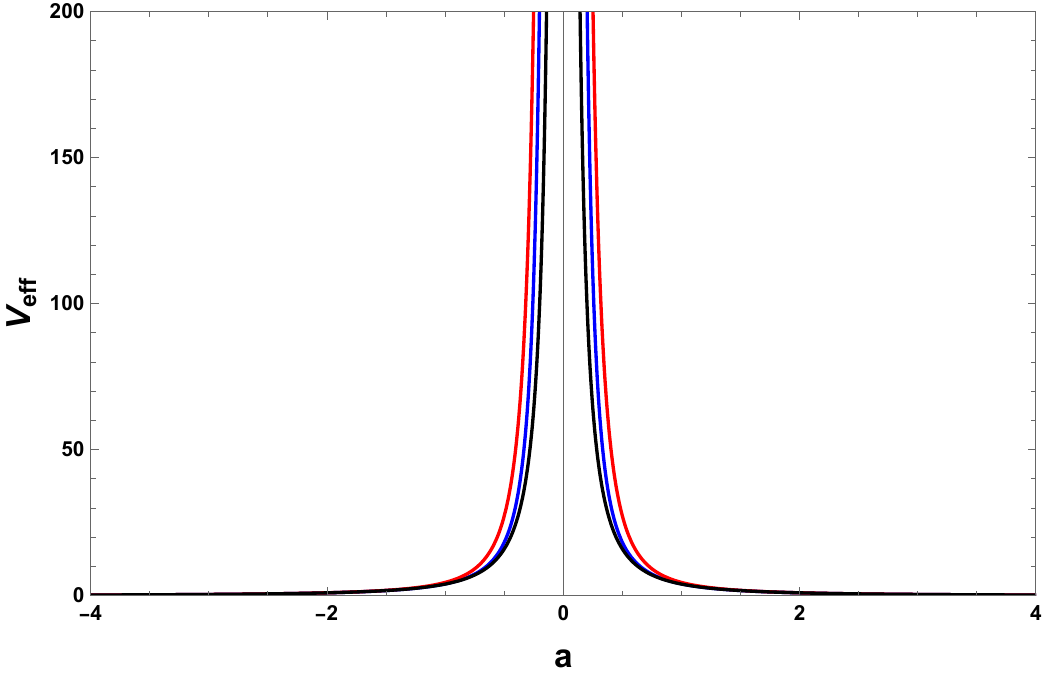}
\includegraphics[angle=0, width=0.40\textwidth]{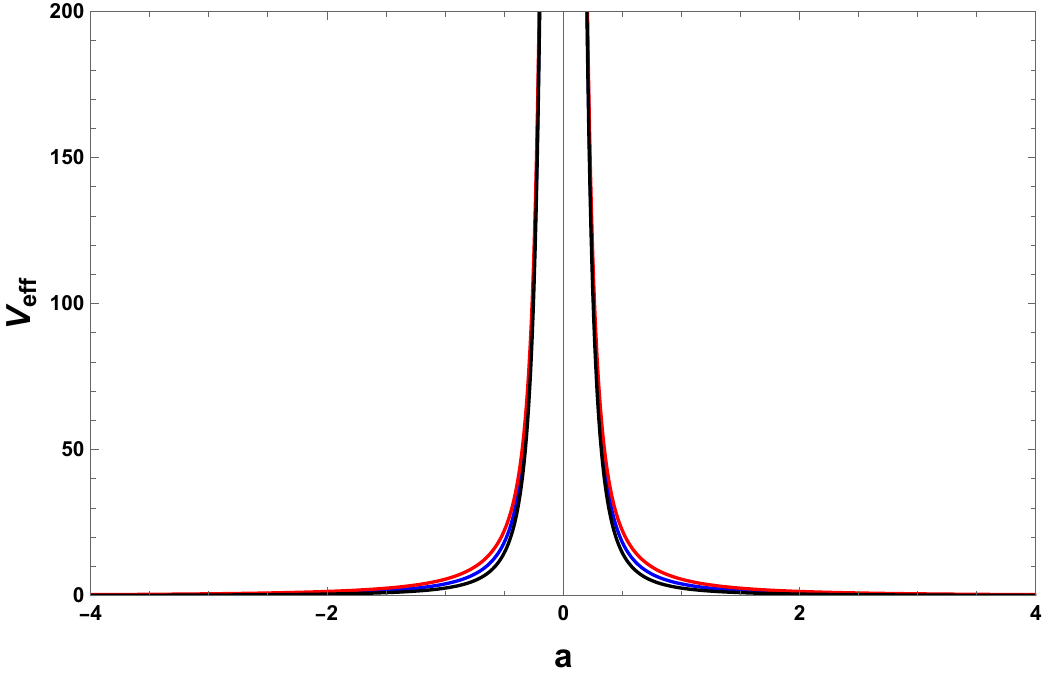}
\caption{Left: Change of effective potential at the photon sphere with $a$ for two sets: Left $\rightarrow$ $C_1 = 1$ and different values of $C_2$ changing. Right $\rightarrow$ $C_2 = 1$ and diferent values of $C_1$. In both graphs, the angular momentum parameter is set to $ L = 2$.}
\label{fig_71}
\end{center}
\end{figure}

\begin{figure}[t!]
\begin{center}
\includegraphics[angle=0, width=0.40\textwidth]{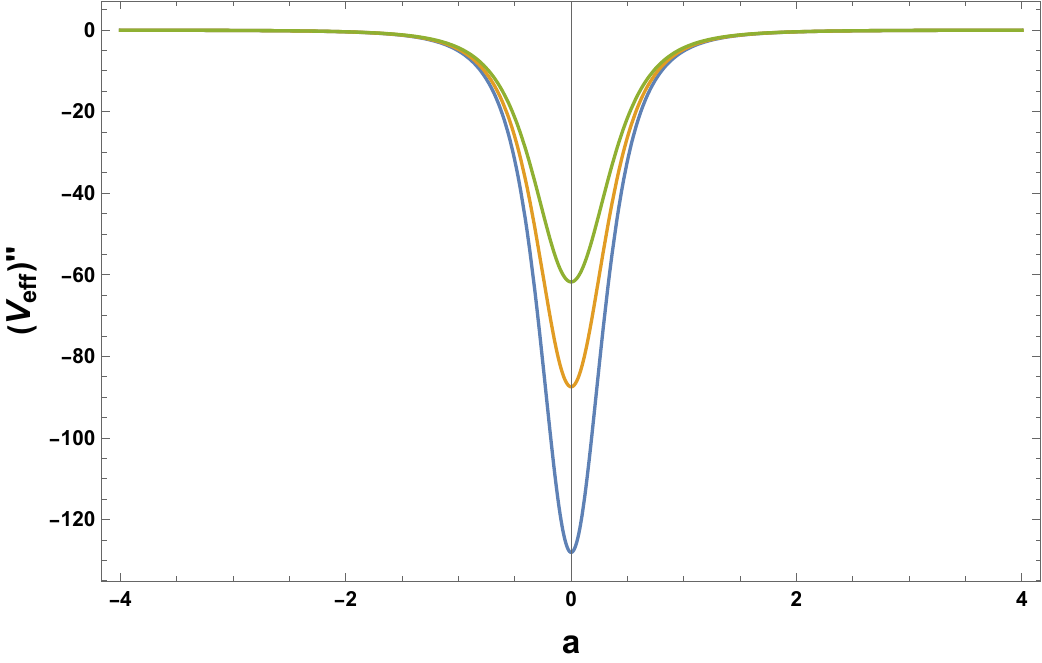}
\caption{Second derivative of the effective potential at the photon sphere for different values of $a$ and varying $C_2$, with the angular momentum parameter fixed at $L = 2$.}
\label{fig_81}
\end{center}
\end{figure}

For a Phantom wormhole \ref{exactG}, this leads to the condition

\begin{equation}
2 r G = (a^{2}+r^{2}) G' ~~ \Rightarrow ~~ \frac{L^2 (C_2 - 2 r_{\rm ph})}{(a^2 + {r_{\rm ph}}^2)^2} = 0, ~~ \Rightarrow ~~ r_{\rm ph} = \frac{C_{2}}{2}.
\end{equation}

A circular photon orbit is stable or unstable depending on the signature of $V_{0}''$ evaluated at $r_{\rm ph}$. The explicit form of $V_{0}''$ evaluated at $r_{\rm ph}$ is

\begin{equation}
V_{0}''(r)\Big|_{r_{\rm ph}} = -\frac{2L^2 \left\lbrace a^2 + \frac{{C_2}^2}{4} \right\rbrace}{(a^2 + {r_{\rm ph}}^2)^3},
\end{equation}

\begin{figure}[t!]
\begin{center}
\includegraphics[angle=0, width=0.40\textwidth]{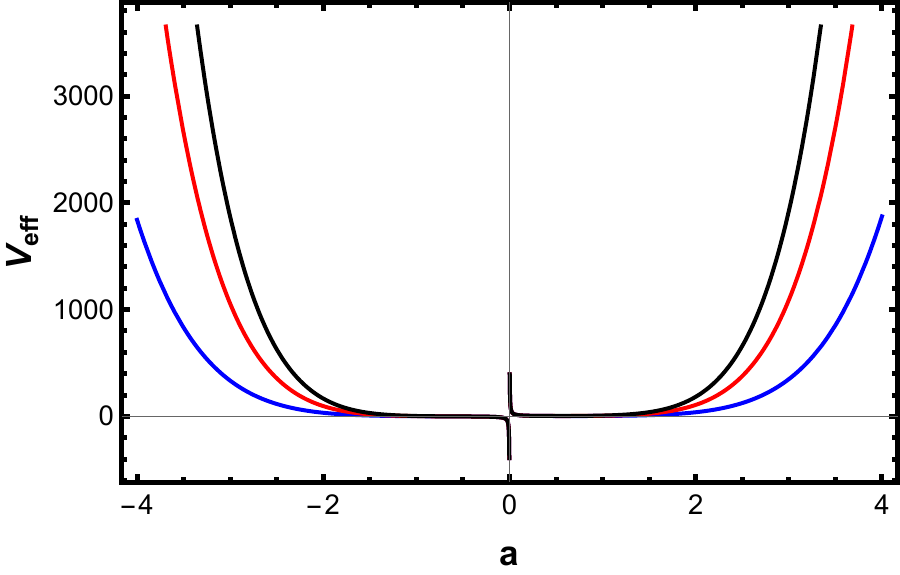}
\includegraphics[angle=0, width=0.40\textwidth]{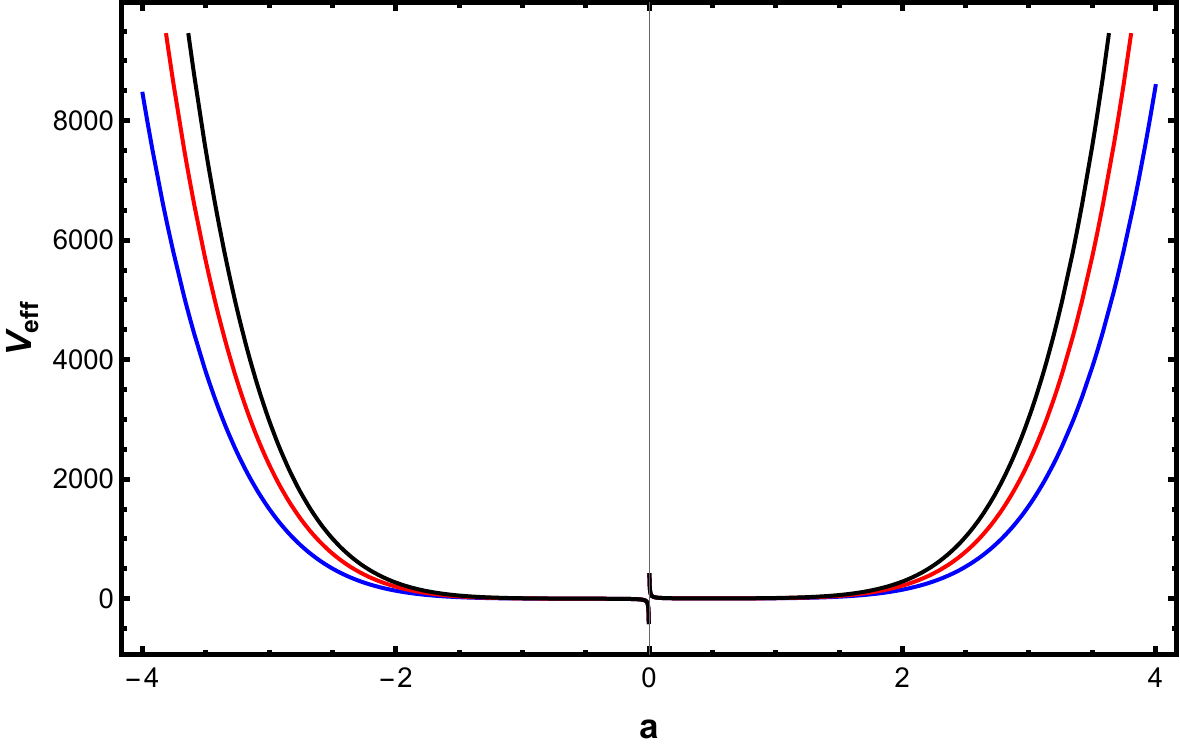}
\caption{Left: Change of effective potential at the photon sphere with $a$ for two sets: Left $\rightarrow$ $C_1 = 1$ and different values of $C_2$ changing. Right $\rightarrow$ $C_2 = 1$ and diferent values of $C_1$. In both graphs, the angular momentum parameter is set to $ L = 2$.}
\label{fig_711}
\end{center}
\end{figure}

\begin{figure}[t!]
\begin{center}
\includegraphics[angle=0, width=0.40\textwidth]{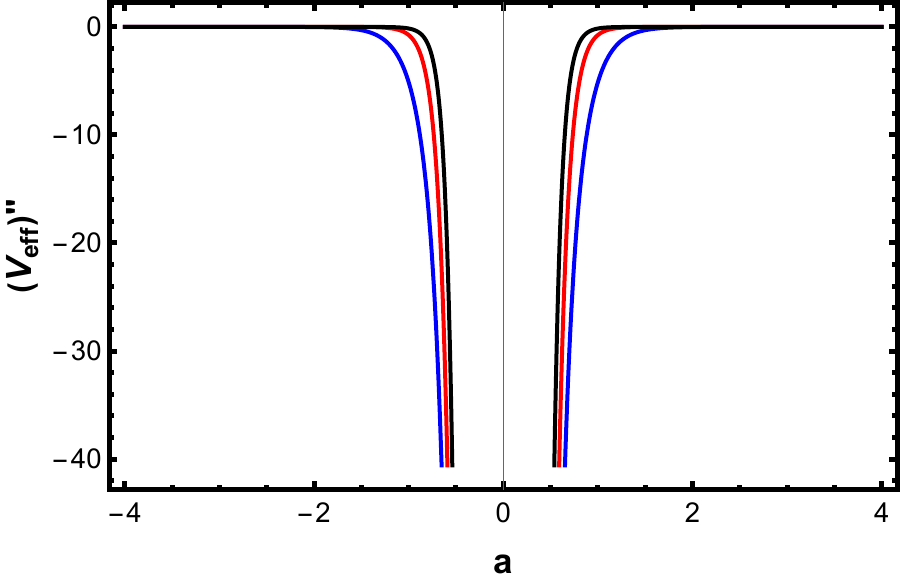}
\caption{Second derivative of the effective potential at the photon sphere for different values of $a$ and varying $C_2$, with the angular momentum parameter fixed at $L = 2$.}
\label{fig_811}
\end{center}
\end{figure}

which is always negative and confirms the existence of unstable photon orbits for a Phanton wormhole spacetime. This behavior is independent of the choice of $a$. We demonstrate this in \ref{fig_71} and \ref{fig_81} by plotting the effective potential and its' second derivative for different values of $a$, taking $L = 2$. Similarly, the photon orbit radius for the Kiselev wormhole metric can be calculated as
\begin{equation}
r_{\rm ph} = -a^3 C_2 p,
\end{equation}

which indicates that a physical photon orbit can only develop if either $C_2$ or $p$ is chosen as negative. The second derivative of the effective potential $V_{0}''$ is evaluated at $r_{\rm ph}$ as

\begin{equation}
V_{0}''(r)\Big|_{r_{\rm ph}} = -\frac{2 L^2 \left(a^6 C_{2}^2 p^2 + a^2\right)}{\left(a^6 C_{2}^2 p^2 + a^2\right)^3},
\end{equation}

which is, again, always negative. This confirms the existence of unstable photon orbits for the Kiselev wormhole spacetime as well. The overall behavior is independent of the choice of $a$ which we demonstrate in \ref{fig_711} and \ref{fig_811}. We plot the effective potential and its' second derivative for different values of $a$, taking $L = 2$. \\

A Lyapunov exponent is connected to small perturbations around the radial null trajectory defined in a metric tensor of a compact object, in this case, the wormhole metrics. We discuss how the effective potential, $V_{0}$, defined in \ref{effpot}, is affected by a small perturbation $\delta r$ close to the photon orbit radius $r = r_{\rm{ph}}$. We derive to an expression for the evolution of the perturbation, given by $\delta r \sim \exp(\lambda t)$, where $\lambda$ is the Lyapunov exponent. This exponent quantifies the rate at which perturbations can grow or decay over time, characterizing the stability of the orbit as well as the object \citep{Mishra:2020jlw, Rahman:2018oso, Cardoso:2017soq, Cardoso:2008bp}. The generic expression is found as

\begin{equation}
\lambda = \sqrt{\frac{G(r_{\rm ph})}{2} \left(\frac{2 G(r_{\rm ph})}{r^2_{\rm ph}} - G''(r_{\rm ph})\right)}.
\end{equation}

\begin{figure}[t!]
\begin{center}
\includegraphics[angle=0, width=0.40\textwidth]{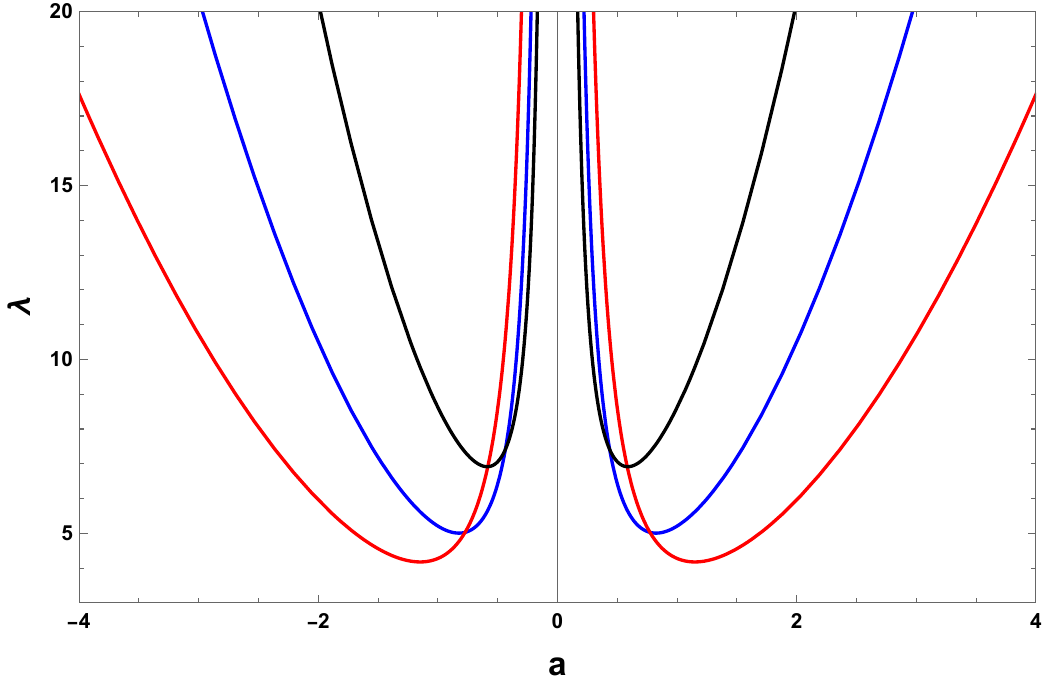}
\includegraphics[angle=0, width=0.40\textwidth]{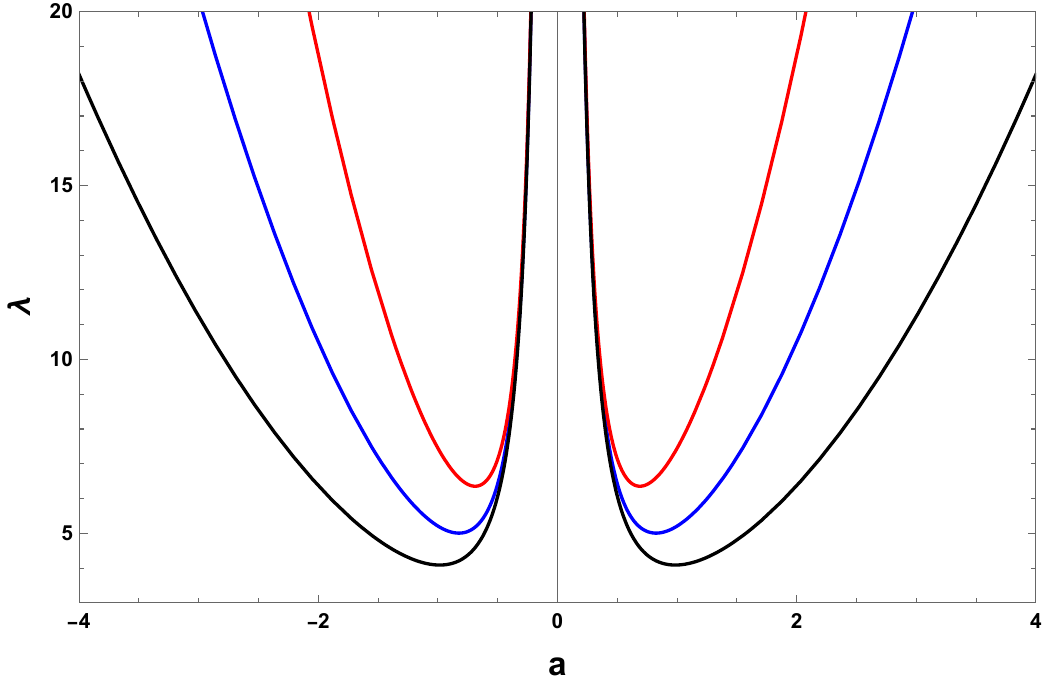}
\caption{Left: Plot of the Lyapunov exponent for a Phantom wormhole as a function of $a$, for $C_{1}=1$ and different values of $C_2$. Right: Plot of the Lyapunov exponent as a function of $a$, for $C_{2}=1$ and different values of $C_1$. }
\label{fig_8}
\end{center}
\end{figure}

The exponent is dependent on metric coefficients, which indicate a correlation of the trajectory with the underlying spacetime geometry. Using \ref{exactG} and the radius of the photon sphere for Phantom wormhole, we calculate the Lyapunov exponent as
\begin{equation}
\lambda = \frac{1}{2} \sqrt{\frac{\left(2 \left(a^3 C_{1} + a\right) + C_{2} \cot^{-1}\left(\frac{2 a}{C_{2}}\right)\right) \left(8 a^5 C_{1} + 2 a^3 \left(C_{1} C_{2}^2 + 4\right) + \left(4 a^2 C_{2} + C_{2}^3\right) \cot^{-1}\left(\frac{2 a}{C_{2}}\right) + 4 a C_{2}^2\right)}{a^4 C_{2}^2}}~.
\end{equation}

\begin{figure}[t!]
\begin{center}
\includegraphics[angle=0, width=0.40\textwidth]{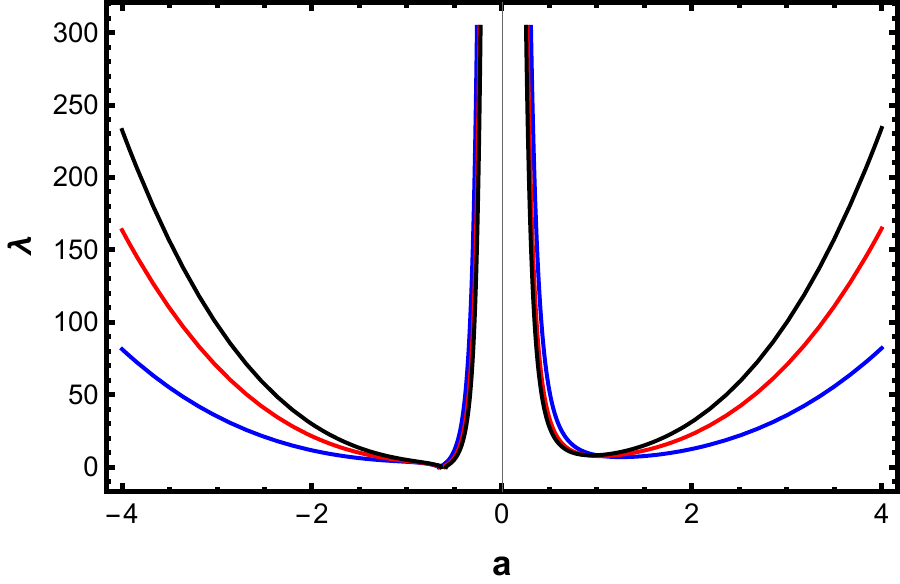}
\includegraphics[angle=0, width=0.40\textwidth]{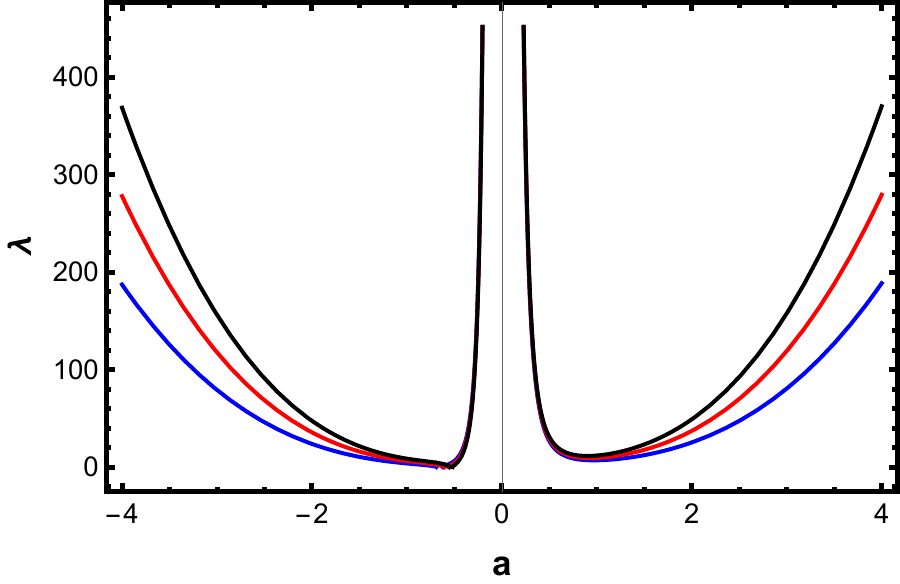}
\caption{Left: Plot of the Lyapunov exponent for a Kiselev wormhole as a function of $a$, for $C_{1}=1$ and different values of $C_2$. Right: Plot of the Lyapunov exponent as a function of $a$, for $C_{2}=1$ and different values of $C_1$. }
\label{fig_80}
\end{center}
\end{figure}

For a Phantom wormhole we plot $\lambda$ in \ref{fig_8} for different values of $a$. Similarly, the Lyapunov exponent for the Kiselev wormhole is shown in \ref{fig_80}. For both of the cases, a positive Lyapunov exponent signifies that nearby trajectories diverge over time, indicating a strong sensitivity to initial conditions. \\

The shadow radius for a compact object refers to the apparent size of the dark shape cast by the object against a backdrop of light, usually from a bright source behind it. One can define a critical impact parameter which corresponds to the boundary between light rays that escape to infinity and those that are captured by the geometry (which in this case is either a one-way or a traversable wormhole). For a Phantom wormhole, the shadow radius $r_{sh}$ is calculated by finding the critical parameter associated with null geodesics \citep{book:338838} for \ref{exactG},

\begin{equation}
\frac{1}{r^{2}_{\textrm{sh}}} = \frac{G(r_{\rm ph})}{(r_{\rm ph}^{2} + a^{2})} ~~\Rightarrow~~ r_{sh} = \sqrt{\frac{2 a^3}{2 a^3 C_{1} + C_{2} \tan^{-1}\left(\frac{C_{2}}{2 a}\right) + 2 a}}~,
\end{equation}

\begin{figure}[t!]
\begin{center}
\includegraphics[angle=0, width=0.40\textwidth]{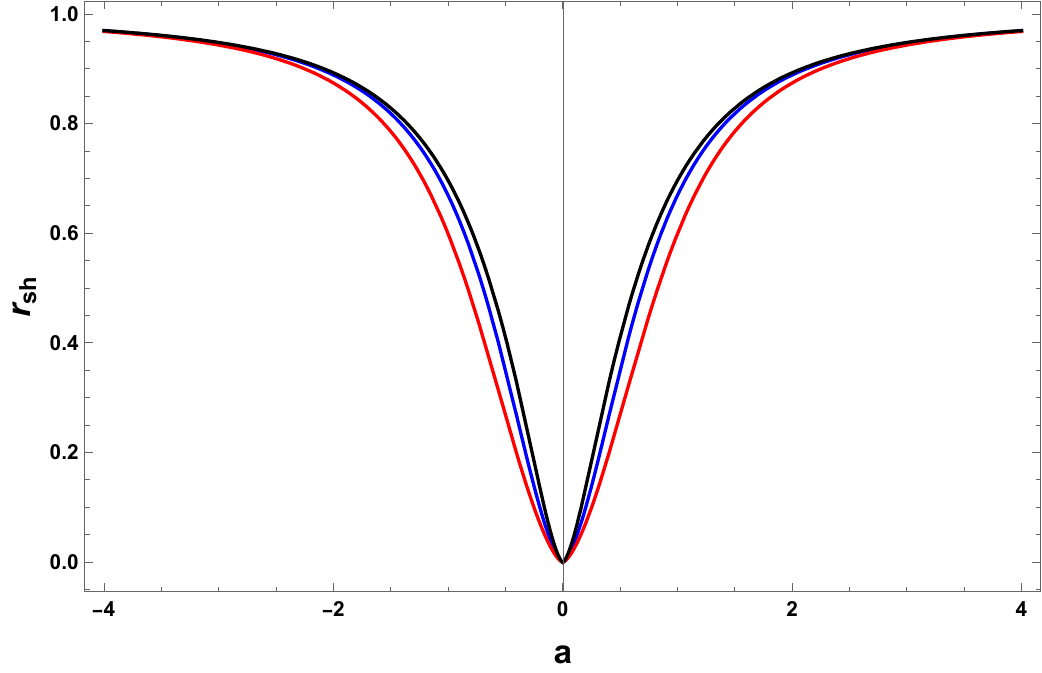}
\includegraphics[angle=0, width=0.40\textwidth]{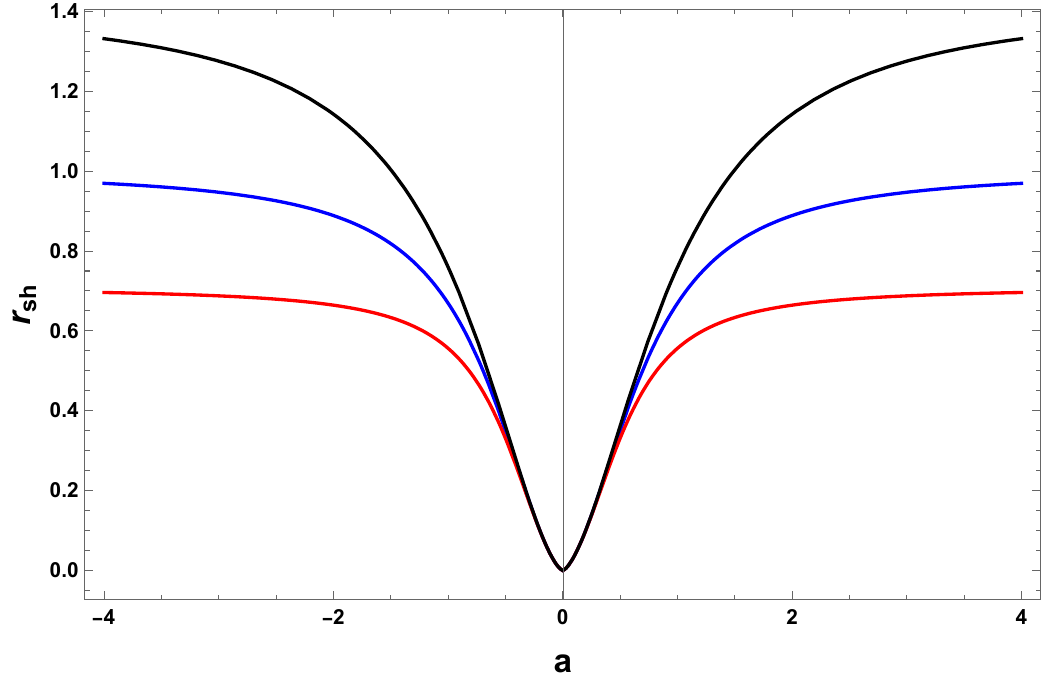}
\caption{Shadow radius of a Phantom wormhole for different values of $a$. Left: $C_{1}$ is fixed at $1$ while different values of $C_2$ are chosen. Right : $C_{2}$ is fixed at $1$ while different values of $C_{1}$ is considered.}
\label{fig_9}
\end{center}
\end{figure}

\begin{figure}[t!]
\begin{center}
\includegraphics[angle=0, width=0.40\textwidth]{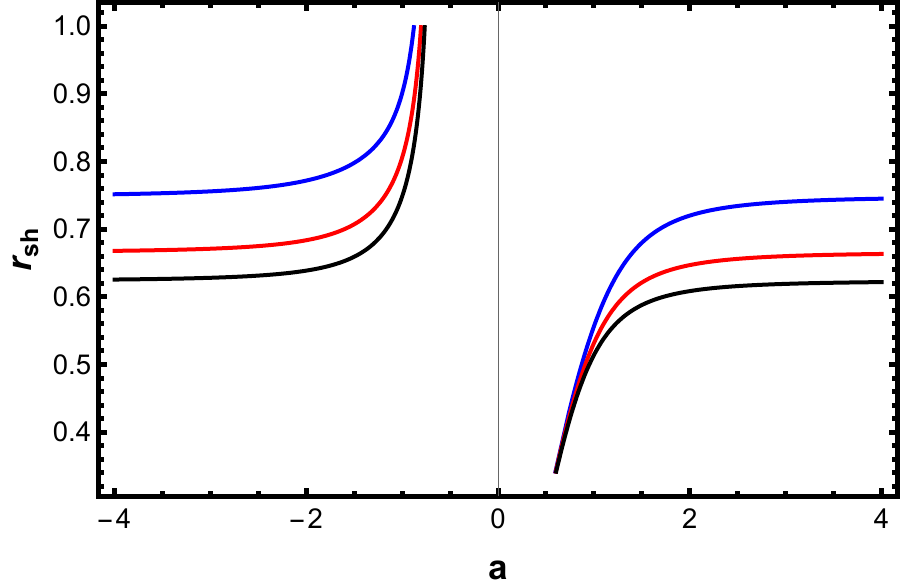}
\includegraphics[angle=0, width=0.40\textwidth]{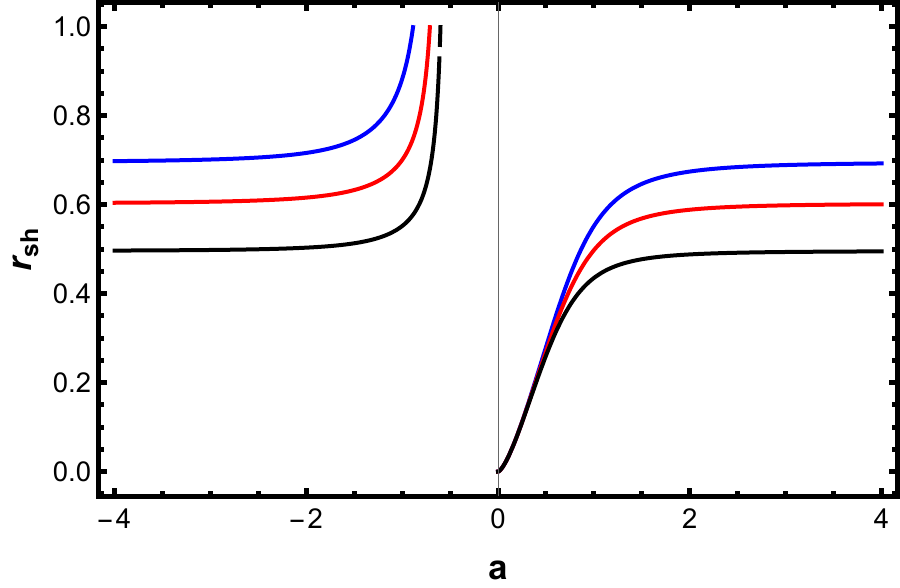}
\caption{Shadow radius of a Kiselev wormhole for different values of $a$ for Left : $C_{1}=1$ and different values of $C_2$ ; Right : $C_{2}=1$ and different values of $C_1$.}
\label{fig_90}
\end{center}
\end{figure}

We plot the shadow radius in \ref{fig_9} as a function of $a$ for different sets of $C_{1}$ and $C_2$. Similarly, the shadow radius for the Kiselev wormhole is found as
\begin{equation}
r_{sh} = \sqrt{\frac{a^3}{a^3 C_{1} p + a^3 C_{2} p \tan^{-1}\left(a^2 C_{2} p \right)+a+m}}~,
\end{equation}
which is plotted in \ref{fig_90}.

Geodesic orbits for massive particles can be described using a timelike worldline for the $\epsilon = -1$ case of \ref{effpot}. The innermost stable circular orbit then corresponds to the condition that the effective potential experienced by the particle must obey $V'_{-1} = 0$ \citep{Cardoso:2008bp}. For the Phantom wormhole, this can be simplified into
\begin{equation}
2a^7 {C_1}r + a {C_2}r^4 + a^5 ({C_2} + 4{C_1}r^3) + a^3 ({C_2}L^2 - 2L^2 r + 2{C_2}r^2 + 2{C_1}r^5) + {C_2}r(a^2 + r^2)\tan^{-1}\left(\frac{r}{a}\right) = 0.
\end{equation}

One can solve this equation numerically, however it is better to proceed with a fixed coordinate value $r = r_{c}$ where a circular orbit is expected to form. At this coordinate value we derive the required angular momentum $L_c$, in terms of $r_{c}$, $a$, $C_1$, and $C_2$. Therefore the innermost circular orbit for a massive particle forms at the coordinate value $r_c$ for which $L_c$ is minimized. Since with a $\tan^{-1}\left(\frac{r_c}{a}\right)$ term, it is non-trivial to proceed, we study the condition within a small neighborhood around $r = 0$. In this limit we approximate $\tan^{-1}\left(\frac{r}{a}\right) \simeq \frac{r}{a}$ and neglect the higher order terms of $r_c$ to write 

\begin{equation}
L_c = \sqrt{\frac{2a^7 C_1 r_c + a^5 C_2}{a^3 (2r_c - C_2)}}.
\end{equation}

This expression can be compared with circular orbits in weak-field GR. In classical physics, the angular momentum per unit mass for a particle with angular velocity $\omega$ is $L_c \sim \omega r_c$. Kepler's third law of planetary motion implies that $\omega^2 \sim \frac{G_N m}{r_c}$, where $m$ is the mass of a central object. Therefore one can write that
\begin{equation}
L_c \sim \sqrt{\frac{G_N m}{r_c}}r_c \Rightarrow L_c \sim \sqrt{m r_c}.
\end{equation}
In comparison with the weak field limit, we note that for a Phantom wormhole, the parametric combination of $   \frac{2a^4 C_1}{C_2}$ has a connection with the corresponding mass function. Similarly, the $L_{c}$ for a Kiselev wormhole is derived as

\begin{equation}
L_c = \frac{\sqrt{-a^5 C_{2} p + a^4 C_{1} p r_c + a m r_c}}{\sqrt{a^3 C_{2} p + r_c}}.
\end{equation}

\section{Conclusion}
The idea of matter holds particular importance in GR. It is the connection between an Einstein tensor and an energy-momentum tensor (both being free of covariant divergence) that fabricates GR out of Riemannian geometry. It can not be denied that the field equations describing this connection between geometry and matter, can produce exotic solutions. Some of these solutions carry space-time points of infinite curvature within their boundary/horizon while some them behave a bit more radically and allow a chance of safe passage through to a distant part of the universe. The genesis of the matter and energy that allows a formation of such exotic objects remains an interesting topic of research. To that end, the present article explores three notions : (i) it checks if a non-trivial scalar deformation of the Schwarzschild solution can exist for a static spherically symmetric GR plus scalar field system ; (ii) it explores the method of regularization of an existing singular solution of GR and the properties of the resulting geometry and (iii) it forms a connection between the formation of wormhole throat to an effective phase transition of the matter field involved, through a spontaneous symmetry breaking. \\ 

\begin{figure}[t!]
\begin{center}
\includegraphics[angle=0, width=0.40\textwidth]{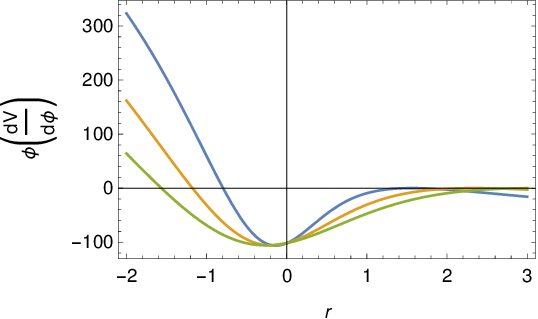}
\includegraphics[angle=0, width=0.40\textwidth]{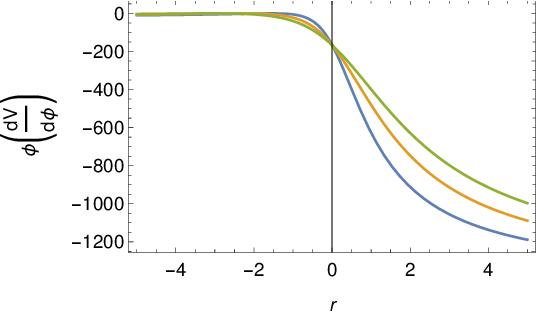}
\caption{$\phi \frac{dV}{d\phi}$ as a function of $r$ for two different profiles of scalar field $\phi(r) = C_{3} \pm 2 \tan^{-1}(r/a)$.}
\label{fig_13}
\end{center}
\end{figure}

For any massive (self-interaction potential $V(\phi)$) scalar field minimally coupled to GR, the condition $\phi \frac{dV}{d\phi} \geq 0$ is usually treated as a requirement such that the self-interaction potential has a local minimum at $\phi = 0$ and no local maximum within the $\phi$-range considered. This condition can be treated as a key to effectively ruling out non-trivial scalar deformations of the Schwarzschild solution. For the present solution it can be seen in \ref{fig_13} that the condition is always violated irrespective of the scalar field profile under consideration. \\

The solutions we explore in this article describe two-way traversable wormholes or, in selected cases, regular black holes. The first solution can be traced back to the regular black hole solution of Bronnikov \citep{Bronnikov_2006}, which is supported by a phantom scalar field (negative kinetic energy). We connect the metric with a second class of solution popular as the Kiselev black hole \citep{Kiselev_2003}. We derive a generalized version of the Kiselev metric. We prove that both of these solutions can be supported by a massive scalar field whose self-interaction is Higgs type, i.e., has a linear combination of quadratic and quartic terms. Moreover, the solutions are consistent if the coefficient of the quadratic term, as in the mass term $M$, is a function of coordinates (in the form of \ref{potential}). The functional form of $M$ is solved from the field equations and it can be seen that the profile carries a signature of the breakdown of $Z_2$ symmetry in the system, around the wormhole throat radius. The solutions do not exhibit any singularity as checked from the curvature scalar profiles which remain finite for all possible coordinate ranges. The behavior of radial null geodesics demonstrates that the metrics represent either of the three: a one-way wormhole, a two-way traversable wormhole or a regular blackhole, depending on the parameter space. We compute and discuss properties of the photon sphere, Lyapunov exponent, shadow radius and the radius of the innermost stable circular orbits for the metrics. It may be an interesting prospect to further generalize or compare these solutions with \textit{dirty} black holes \citep{PhysRevD.46.2445, PhysRevD.48.583, PhysRevD.88.041502}, associated with non-trivial stress-energy tensors \citep{Simpson_2019}. This is usually done through the analysis of a derived metric tensor in generalized tortoise coordinates. A subsequent derivation of the Regge-Wheeler equations follows for spin zero, spin one, and/or spin two particles coupled to gravity. This will be discussed in a follow-up article.

\section*{Acknowledgement}

Soumya Chakrabarti acknowledges the IUCAA for providing facility and support under the visiting associateship program. Acknowledgment is also given to Vellore Institute of Technology for the financial support through its Seed Grant (No. SG20230027), the year 2023.

\bibliography{references}

\begin{thebibliography}{83}%
\makeatletter
\providecommand \@ifxundefined [1]{%
 \@ifx{#1\undefined}
}%
\providecommand \@ifnum [1]{%
 \ifnum #1\expandafter \@firstoftwo
 \else \expandafter \@secondoftwo
 \fi
}%
\providecommand \@ifx [1]{%
 \ifx #1\expandafter \@firstoftwo
 \else \expandafter \@secondoftwo
 \fi
}%
\providecommand \natexlab [1]{#1}%
\providecommand \enquote  [1]{``#1''}%
\providecommand \bibnamefont  [1]{#1}%
\providecommand \bibfnamefont [1]{#1}%
\providecommand \citenamefont [1]{#1}%
\providecommand \href@noop [0]{\@secondoftwo}%
\providecommand \href [0]{\begingroup \@sanitize@url \@href}%
\providecommand \@href[1]{\@@startlink{#1}\@@href}%
\providecommand \@@href[1]{\endgroup#1\@@endlink}%
\providecommand \@sanitize@url [0]{\catcode `\\12\catcode `\$12\catcode
  `\&12\catcode `\#12\catcode `\^12\catcode `\_12\catcode `\%12\relax}%
\providecommand \@@startlink[1]{}%
\providecommand \@@endlink[0]{}%
\providecommand \url  [0]{\begingroup\@sanitize@url \@url }%
\providecommand \@url [1]{\endgroup\@href {#1}{\urlprefix }}%
\providecommand \urlprefix  [0]{URL }%
\providecommand \Eprint [0]{\href }%
\providecommand \doibase [0]{http://dx.doi.org/}%
\providecommand \selectlanguage [0]{\@gobble}%
\providecommand \bibinfo  [0]{\@secondoftwo}%
\providecommand \bibfield  [0]{\@secondoftwo}%
\providecommand \translation [1]{[#1]}%
\providecommand \BibitemOpen [0]{}%
\providecommand \bibitemStop [0]{}%
\providecommand \bibitemNoStop [0]{.\EOS\space}%
\providecommand \EOS [0]{\spacefactor3000\relax}%
\providecommand \BibitemShut  [1]{\csname bibitem#1\endcsname}%
\let\auto@bib@innerbib\@empty
\bibitem [{\citenamefont {{Schwarzschild}}(1916)}]{1999physics...5030S}%
  \BibitemOpen
  \bibfield  {author} {\bibinfo {author} {\bibfnamefont {K.}~\bibnamefont
  {{Schwarzschild}}},\ }\href {\doibase 10.48550/arXiv.physics/9905030}
  {\bibfield  {journal} {\bibinfo  {journal} {Sitzungsber. Preuss. Akad. Wiss.
  Phys. Math. Kl.,}\ ,\ \bibinfo {eid} {physics/9905030}} (\bibinfo {year}
  {1916})},\ \Eprint {http://arxiv.org/abs/physics/9905030}
  {arXiv:physics/9905030 [physics.hist-ph]} \BibitemShut {NoStop}%
\bibitem [{\citenamefont {{Hilbert}}(1917)}]{hilbert}%
  \BibitemOpen
  \bibfield  {author} {\bibinfo {author} {\bibfnamefont {D.}~\bibnamefont
  {{Hilbert}}},\ }\href@noop {} {\bibfield  {journal} {\bibinfo  {journal}
  {Nachr. Ges. Wiss. Math. Phys. Kl. (G¨ottingen)}\ ,\ \bibinfo {pages} {53}}
  (\bibinfo {year} {1917})}\BibitemShut {NoStop}%
\bibitem [{\citenamefont {{Weyl}}(1917)}]{weyl}%
  \BibitemOpen
  \bibfield  {author} {\bibinfo {author} {\bibfnamefont {H.}~\bibnamefont
  {{Weyl}}},\ }\href@noop {} {\bibfield  {journal} {\bibinfo  {journal} {Ann.
  d. Phys.}\ }\textbf {\bibinfo {volume} {54}},\ \bibinfo {pages} {117}
  (\bibinfo {year} {1917})}\BibitemShut {NoStop}%
\bibitem [{\citenamefont {{Jebsen}}(1921)}]{jebsen}%
  \BibitemOpen
  \bibfield  {author} {\bibinfo {author} {\bibfnamefont {J.~T.}\ \bibnamefont
  {{Jebsen}}},\ }\href@noop {} {\bibfield  {journal} {\bibinfo  {journal} {Ark.
  Mat. Ast. Fys. (Stockholm)}\ }\textbf {\bibinfo {volume} {15}},\ \bibinfo
  {pages} {18} (\bibinfo {year} {1921})}\BibitemShut {NoStop}%
\bibitem [{\citenamefont {{Birkhoff}}(1923)}]{birkhoff}%
  \BibitemOpen
  \bibfield  {author} {\bibinfo {author} {\bibfnamefont {G.~D.}\ \bibnamefont
  {{Birkhoff}}},\ }\href@noop {} {\bibfield  {journal} {\bibinfo  {journal}
  {Relativity and Modern Physics, Harvard University Press}\ }\textbf {\bibinfo
  {volume} {253}},\ \bibinfo {pages} {18} (\bibinfo {year} {1923})}\BibitemShut
  {NoStop}%
\bibitem [{\citenamefont {{Sato}}(1995)}]{nh1}%
  \BibitemOpen
  \bibfield  {author} {\bibinfo {author} {\bibfnamefont {K.}~\bibnamefont
  {{Sato}}},\ }\href@noop {} {\bibfield  {journal} {\bibinfo  {journal}
  {Journal of Astrophysics and Astronomy Supplement}\ }\textbf {\bibinfo
  {volume} {16}},\ \bibinfo {pages} {37} (\bibinfo {year} {1995})}\BibitemShut
  {NoStop}%
\bibitem [{\citenamefont {Bechmann}\ and\ \citenamefont
  {Lechtenfeld}(1995)}]{nh2}%
  \BibitemOpen
  \bibfield  {author} {\bibinfo {author} {\bibfnamefont {O.}~\bibnamefont
  {Bechmann}}\ and\ \bibinfo {author} {\bibfnamefont {O.}~\bibnamefont
  {Lechtenfeld}},\ }\href {\doibase 10.1088/0264-9381/12/6/013} {\bibfield
  {journal} {\bibinfo  {journal} {Class. Quant. Grav.}\ }\textbf {\bibinfo
  {volume} {12}},\ \bibinfo {pages} {1473} (\bibinfo {year} {1995})},\ \Eprint
  {http://arxiv.org/abs/gr-qc/9502011} {arXiv:gr-qc/9502011} \BibitemShut
  {NoStop}%
\bibitem [{\citenamefont {Bekenstein}(1996)}]{nh3}%
  \BibitemOpen
  \bibfield  {author} {\bibinfo {author} {\bibfnamefont {J.~D.}\ \bibnamefont
  {Bekenstein}},\ }in\ \href@noop {} {\emph {\bibinfo {booktitle} {{2nd
  International Sakharov Conference on Physics}}}}\ (\bibinfo {year} {1996})\
  pp.\ \bibinfo {pages} {216--219},\ \Eprint
  {http://arxiv.org/abs/gr-qc/9605059} {arXiv:gr-qc/9605059} \BibitemShut
  {NoStop}%
\bibitem [{\citenamefont {Herdeiro}\ and\ \citenamefont {Radu}(2015)}]{nh4}%
  \BibitemOpen
  \bibfield  {author} {\bibinfo {author} {\bibfnamefont {C.~A.~R.}\
  \bibnamefont {Herdeiro}}\ and\ \bibinfo {author} {\bibfnamefont
  {E.}~\bibnamefont {Radu}},\ }\href {\doibase 10.1142/S0218271815420146}
  {\bibfield  {journal} {\bibinfo  {journal} {International Journal of Modern
  Physics D}\ }\textbf {\bibinfo {volume} {24}},\ \bibinfo {pages} {1542014}
  (\bibinfo {year} {2015})}\BibitemShut {NoStop}%
\bibitem [{\citenamefont {Barceló}\ \emph {et~al.}(2019)\citenamefont
  {Barceló}, \citenamefont {Carballo-Rubio},\ and\ \citenamefont
  {Liberati}}]{nh5}%
  \BibitemOpen
  \bibfield  {author} {\bibinfo {author} {\bibfnamefont {C.}~\bibnamefont
  {Barceló}}, \bibinfo {author} {\bibfnamefont {R.}~\bibnamefont
  {Carballo-Rubio}}, \ and\ \bibinfo {author} {\bibfnamefont {S.}~\bibnamefont
  {Liberati}},\ }\href {\doibase 10.1088/1361-6382/ab23b6} {\bibfield
  {journal} {\bibinfo  {journal} {Classical and Quantum Gravity}\ }\textbf
  {\bibinfo {volume} {36}},\ \bibinfo {pages} {13LT01} (\bibinfo {year}
  {2019})}\BibitemShut {NoStop}%
\bibitem [{\citenamefont {Perlmutter}\ \emph {et~al.}(1997)\citenamefont
  {Perlmutter} \emph {et~al.}}]{de1}%
  \BibitemOpen
  \bibfield  {author} {\bibinfo {author} {\bibfnamefont {S.}~\bibnamefont
  {Perlmutter}} \emph {et~al.} (\bibinfo {collaboration} {Supernova Cosmology
  Project}),\ }\href@noop {} {\bibfield  {journal} {\bibinfo  {journal} {Bull.
  Am. Astron. Soc.}\ }\textbf {\bibinfo {volume} {29}},\ \bibinfo {pages}
  {1351} (\bibinfo {year} {1997})},\ \Eprint
  {http://arxiv.org/abs/astro-ph/9812473} {arXiv:astro-ph/9812473} \BibitemShut
  {NoStop}%
\bibitem [{\citenamefont {Riess}\ \emph {et~al.}(1998)\citenamefont {Riess},
  \citenamefont {Filippenko}, \citenamefont {Challis}, \citenamefont
  {Clocchiatti}, \citenamefont {Diercks}, \citenamefont {Garnavich},
  \citenamefont {Gilliland}, \citenamefont {Hogan}, \citenamefont {Jha},
  \citenamefont {Kirshner}, \citenamefont {Leibundgut}, \citenamefont
  {Phillips}, \citenamefont {Reiss}, \citenamefont {Schmidt}, \citenamefont
  {Schommer}, \citenamefont {Smith}, \citenamefont {Spyromilio}, \citenamefont
  {Stubbs}, \citenamefont {Suntzeff},\ and\ \citenamefont {Tonry}}]{de2}%
  \BibitemOpen
  \bibfield  {author} {\bibinfo {author} {\bibfnamefont {A.~G.}\ \bibnamefont
  {Riess}}, \bibinfo {author} {\bibfnamefont {A.~V.}\ \bibnamefont
  {Filippenko}}, \bibinfo {author} {\bibfnamefont {P.}~\bibnamefont {Challis}},
  \bibinfo {author} {\bibfnamefont {A.}~\bibnamefont {Clocchiatti}}, \bibinfo
  {author} {\bibfnamefont {A.}~\bibnamefont {Diercks}}, \bibinfo {author}
  {\bibfnamefont {P.~M.}\ \bibnamefont {Garnavich}}, \bibinfo {author}
  {\bibfnamefont {R.~L.}\ \bibnamefont {Gilliland}}, \bibinfo {author}
  {\bibfnamefont {C.~J.}\ \bibnamefont {Hogan}}, \bibinfo {author}
  {\bibfnamefont {S.}~\bibnamefont {Jha}}, \bibinfo {author} {\bibfnamefont
  {R.~P.}\ \bibnamefont {Kirshner}}, \bibinfo {author} {\bibfnamefont
  {B.}~\bibnamefont {Leibundgut}}, \bibinfo {author} {\bibfnamefont {M.~M.}\
  \bibnamefont {Phillips}}, \bibinfo {author} {\bibfnamefont {D.}~\bibnamefont
  {Reiss}}, \bibinfo {author} {\bibfnamefont {B.~P.}\ \bibnamefont {Schmidt}},
  \bibinfo {author} {\bibfnamefont {R.~A.}\ \bibnamefont {Schommer}}, \bibinfo
  {author} {\bibfnamefont {R.~C.}\ \bibnamefont {Smith}}, \bibinfo {author}
  {\bibfnamefont {J.}~\bibnamefont {Spyromilio}}, \bibinfo {author}
  {\bibfnamefont {C.}~\bibnamefont {Stubbs}}, \bibinfo {author} {\bibfnamefont
  {N.~B.}\ \bibnamefont {Suntzeff}}, \ and\ \bibinfo {author} {\bibfnamefont
  {J.}~\bibnamefont {Tonry}},\ }\href {\doibase 10.1086/300499} {\bibfield
  {journal} {\bibinfo  {journal} {The Astronomical Journal}\ }\textbf {\bibinfo
  {volume} {116}},\ \bibinfo {pages} {1009} (\bibinfo {year}
  {1998})}\BibitemShut {NoStop}%
\bibitem [{\citenamefont {Melchiorri}\ \emph {et~al.}(2000)\citenamefont
  {Melchiorri}, \citenamefont {Ade}, \citenamefont {de~Bernardis},
  \citenamefont {Bock}, \citenamefont {Borrill}, \citenamefont {Boscaleri},
  \citenamefont {Crill}, \citenamefont {Troia}, \citenamefont {Farese},
  \citenamefont {Ferreira}, \citenamefont {Ganga}, \citenamefont {de~Gasperis},
  \citenamefont {Giacometti}, \citenamefont {Hristov}, \citenamefont {Jaffe},
  \citenamefont {Lange}, \citenamefont {Masi}, \citenamefont {Mauskopf},
  \citenamefont {Miglio}, \citenamefont {Netterfield}, \citenamefont {Pascale},
  \citenamefont {Piacentini}, \citenamefont {Romeo}, \citenamefont {Ruhl},\
  and\ \citenamefont {Vittorio}}]{de3}%
  \BibitemOpen
  \bibfield  {author} {\bibinfo {author} {\bibfnamefont {A.}~\bibnamefont
  {Melchiorri}}, \bibinfo {author} {\bibfnamefont {P.~A.~R.}\ \bibnamefont
  {Ade}}, \bibinfo {author} {\bibfnamefont {P.}~\bibnamefont {de~Bernardis}},
  \bibinfo {author} {\bibfnamefont {J.~J.}\ \bibnamefont {Bock}}, \bibinfo
  {author} {\bibfnamefont {J.}~\bibnamefont {Borrill}}, \bibinfo {author}
  {\bibfnamefont {A.}~\bibnamefont {Boscaleri}}, \bibinfo {author}
  {\bibfnamefont {B.~P.}\ \bibnamefont {Crill}}, \bibinfo {author}
  {\bibfnamefont {G.~D.}\ \bibnamefont {Troia}}, \bibinfo {author}
  {\bibfnamefont {P.}~\bibnamefont {Farese}}, \bibinfo {author} {\bibfnamefont
  {P.~G.}\ \bibnamefont {Ferreira}}, \bibinfo {author} {\bibfnamefont
  {K.}~\bibnamefont {Ganga}}, \bibinfo {author} {\bibfnamefont
  {G.}~\bibnamefont {de~Gasperis}}, \bibinfo {author} {\bibfnamefont
  {M.}~\bibnamefont {Giacometti}}, \bibinfo {author} {\bibfnamefont {V.~V.}\
  \bibnamefont {Hristov}}, \bibinfo {author} {\bibfnamefont {A.~H.}\
  \bibnamefont {Jaffe}}, \bibinfo {author} {\bibfnamefont {A.~E.}\ \bibnamefont
  {Lange}}, \bibinfo {author} {\bibfnamefont {S.}~\bibnamefont {Masi}},
  \bibinfo {author} {\bibfnamefont {P.~D.}\ \bibnamefont {Mauskopf}}, \bibinfo
  {author} {\bibfnamefont {L.}~\bibnamefont {Miglio}}, \bibinfo {author}
  {\bibfnamefont {C.~B.}\ \bibnamefont {Netterfield}}, \bibinfo {author}
  {\bibfnamefont {E.}~\bibnamefont {Pascale}}, \bibinfo {author} {\bibfnamefont
  {F.}~\bibnamefont {Piacentini}}, \bibinfo {author} {\bibfnamefont
  {G.}~\bibnamefont {Romeo}}, \bibinfo {author} {\bibfnamefont {J.~E.}\
  \bibnamefont {Ruhl}}, \ and\ \bibinfo {author} {\bibfnamefont
  {N.}~\bibnamefont {Vittorio}},\ }\href {\doibase 10.1086/312744} {\bibfield
  {journal} {\bibinfo  {journal} {The Astrophysical Journal}\ }\textbf
  {\bibinfo {volume} {536}},\ \bibinfo {pages} {L63} (\bibinfo {year}
  {2000})}\BibitemShut {NoStop}%
\bibitem [{\citenamefont {Lange}\ \emph {et~al.}(2001)\citenamefont {Lange},
  \citenamefont {Ade}, \citenamefont {Bock}, \citenamefont {Bond},
  \citenamefont {Borrill}, \citenamefont {Boscaleri}, \citenamefont {Coble},
  \citenamefont {Crill}, \citenamefont {de~Bernardis}, \citenamefont {Farese},
  \citenamefont {Ferreira}, \citenamefont {Ganga}, \citenamefont {Giacometti},
  \citenamefont {Hivon}, \citenamefont {Hristov}, \citenamefont {Iacoangeli},
  \citenamefont {Jaffe}, \citenamefont {Martinis}, \citenamefont {Masi},
  \citenamefont {Mauskopf}, \citenamefont {Melchiorri}, \citenamefont
  {Montroy}, \citenamefont {Netterfield}, \citenamefont {Pascale},
  \citenamefont {Piacentini}, \citenamefont {Pogosyan}, \citenamefont {Prunet},
  \citenamefont {Rao}, \citenamefont {Romeo}, \citenamefont {Ruhl},
  \citenamefont {Scaramuzzi},\ and\ \citenamefont {Sforna}}]{de4}%
  \BibitemOpen
  \bibfield  {author} {\bibinfo {author} {\bibfnamefont {A.~E.}\ \bibnamefont
  {Lange}}, \bibinfo {author} {\bibfnamefont {P.~A.~R.}\ \bibnamefont {Ade}},
  \bibinfo {author} {\bibfnamefont {J.~J.}\ \bibnamefont {Bock}}, \bibinfo
  {author} {\bibfnamefont {J.~R.}\ \bibnamefont {Bond}}, \bibinfo {author}
  {\bibfnamefont {J.}~\bibnamefont {Borrill}}, \bibinfo {author} {\bibfnamefont
  {A.}~\bibnamefont {Boscaleri}}, \bibinfo {author} {\bibfnamefont
  {K.}~\bibnamefont {Coble}}, \bibinfo {author} {\bibfnamefont {B.~P.}\
  \bibnamefont {Crill}}, \bibinfo {author} {\bibfnamefont {P.}~\bibnamefont
  {de~Bernardis}}, \bibinfo {author} {\bibfnamefont {P.}~\bibnamefont
  {Farese}}, \bibinfo {author} {\bibfnamefont {P.}~\bibnamefont {Ferreira}},
  \bibinfo {author} {\bibfnamefont {K.}~\bibnamefont {Ganga}}, \bibinfo
  {author} {\bibfnamefont {M.}~\bibnamefont {Giacometti}}, \bibinfo {author}
  {\bibfnamefont {E.}~\bibnamefont {Hivon}}, \bibinfo {author} {\bibfnamefont
  {V.~V.}\ \bibnamefont {Hristov}}, \bibinfo {author} {\bibfnamefont
  {A.}~\bibnamefont {Iacoangeli}}, \bibinfo {author} {\bibfnamefont {A.~H.}\
  \bibnamefont {Jaffe}}, \bibinfo {author} {\bibfnamefont {L.}~\bibnamefont
  {Martinis}}, \bibinfo {author} {\bibfnamefont {S.}~\bibnamefont {Masi}},
  \bibinfo {author} {\bibfnamefont {P.~D.}\ \bibnamefont {Mauskopf}}, \bibinfo
  {author} {\bibfnamefont {A.}~\bibnamefont {Melchiorri}}, \bibinfo {author}
  {\bibfnamefont {T.}~\bibnamefont {Montroy}}, \bibinfo {author} {\bibfnamefont
  {C.~B.}\ \bibnamefont {Netterfield}}, \bibinfo {author} {\bibfnamefont
  {E.}~\bibnamefont {Pascale}}, \bibinfo {author} {\bibfnamefont
  {F.}~\bibnamefont {Piacentini}}, \bibinfo {author} {\bibfnamefont
  {D.}~\bibnamefont {Pogosyan}}, \bibinfo {author} {\bibfnamefont
  {S.}~\bibnamefont {Prunet}}, \bibinfo {author} {\bibfnamefont
  {S.}~\bibnamefont {Rao}}, \bibinfo {author} {\bibfnamefont {G.}~\bibnamefont
  {Romeo}}, \bibinfo {author} {\bibfnamefont {J.~E.}\ \bibnamefont {Ruhl}},
  \bibinfo {author} {\bibfnamefont {F.}~\bibnamefont {Scaramuzzi}}, \ and\
  \bibinfo {author} {\bibfnamefont {D.}~\bibnamefont {Sforna}},\ }\href
  {\doibase 10.1103/PhysRevD.63.042001} {\bibfield  {journal} {\bibinfo
  {journal} {Phys. Rev. D}\ }\textbf {\bibinfo {volume} {63}},\ \bibinfo
  {pages} {042001} (\bibinfo {year} {2001})}\BibitemShut {NoStop}%
\bibitem [{\citenamefont {Padmanabhan}\ and\ \citenamefont
  {Choudhury}(2003)}]{de5}%
  \BibitemOpen
  \bibfield  {author} {\bibinfo {author} {\bibfnamefont {T.}~\bibnamefont
  {Padmanabhan}}\ and\ \bibinfo {author} {\bibfnamefont {T.~R.}\ \bibnamefont
  {Choudhury}},\ }\href {\doibase 10.1046/j.1365-8711.2003.06873.x} {\bibfield
  {journal} {\bibinfo  {journal} {Monthly Notices of the Royal Astronomical
  Society}\ }\textbf {\bibinfo {volume} {344}},\ \bibinfo {pages} {823}
  (\bibinfo {year} {2003})},\ \Eprint
  {http://arxiv.org/abs/https://academic.oup.com/mnras/article-pdf/344/3/823/3304941/344-3-823.pdf}
  {https://academic.oup.com/mnras/article-pdf/344/3/823/3304941/344-3-823.pdf}
  \BibitemShut {NoStop}%
\bibitem [{\citenamefont {Zlatev}\ \emph {et~al.}(1999)\citenamefont {Zlatev},
  \citenamefont {Wang},\ and\ \citenamefont {Steinhardt}}]{scalar1}%
  \BibitemOpen
  \bibfield  {author} {\bibinfo {author} {\bibfnamefont {I.}~\bibnamefont
  {Zlatev}}, \bibinfo {author} {\bibfnamefont {L.}~\bibnamefont {Wang}}, \ and\
  \bibinfo {author} {\bibfnamefont {P.~J.}\ \bibnamefont {Steinhardt}},\ }\href
  {\doibase 10.1103/PhysRevLett.82.896} {\bibfield  {journal} {\bibinfo
  {journal} {Phys. Rev. Lett.}\ }\textbf {\bibinfo {volume} {82}},\ \bibinfo
  {pages} {896} (\bibinfo {year} {1999})}\BibitemShut {NoStop}%
\bibitem [{\citenamefont {Sahni}\ and\ \citenamefont
  {Starobinsky}(2000)}]{scalar2}%
  \BibitemOpen
  \bibfield  {author} {\bibinfo {author} {\bibfnamefont {V.}~\bibnamefont
  {Sahni}}\ and\ \bibinfo {author} {\bibfnamefont {A.}~\bibnamefont
  {Starobinsky}},\ }\href {\doibase 10.1142/S0218271800000542} {\bibfield
  {journal} {\bibinfo  {journal} {International Journal of Modern Physics D}\
  }\textbf {\bibinfo {volume} {09}},\ \bibinfo {pages} {373} (\bibinfo {year}
  {2000})}\BibitemShut {NoStop}%
\bibitem [{\citenamefont {Adelberger}\ \emph {et~al.}(2003)\citenamefont
  {Adelberger}, \citenamefont {Heckel},\ and\ \citenamefont
  {Nelson}}]{Adelberger:2003zx}%
  \BibitemOpen
  \bibfield  {author} {\bibinfo {author} {\bibfnamefont {E.~G.}\ \bibnamefont
  {Adelberger}}, \bibinfo {author} {\bibfnamefont {B.~R.}\ \bibnamefont
  {Heckel}}, \ and\ \bibinfo {author} {\bibfnamefont {A.~E.}\ \bibnamefont
  {Nelson}},\ }\href {\doibase 10.1146/annurev.nucl.53.041002.110503}
  {\bibfield  {journal} {\bibinfo  {journal} {Ann. Rev. Nucl. Part. Sci.}\
  }\textbf {\bibinfo {volume} {53}},\ \bibinfo {pages} {77} (\bibinfo {year}
  {2003})},\ \Eprint {http://arxiv.org/abs/hep-ph/0307284}
  {arXiv:hep-ph/0307284} \BibitemShut {NoStop}%
\bibitem [{\citenamefont {Frieman}\ \emph {et~al.}(1995)\citenamefont
  {Frieman}, \citenamefont {Hill}, \citenamefont {Stebbins},\ and\
  \citenamefont {Waga}}]{PhysRevLett.75.2077}%
  \BibitemOpen
  \bibfield  {author} {\bibinfo {author} {\bibfnamefont {J.~A.}\ \bibnamefont
  {Frieman}}, \bibinfo {author} {\bibfnamefont {C.~T.}\ \bibnamefont {Hill}},
  \bibinfo {author} {\bibfnamefont {A.}~\bibnamefont {Stebbins}}, \ and\
  \bibinfo {author} {\bibfnamefont {I.}~\bibnamefont {Waga}},\ }\href {\doibase
  10.1103/PhysRevLett.75.2077} {\bibfield  {journal} {\bibinfo  {journal}
  {Phys. Rev. Lett.}\ }\textbf {\bibinfo {volume} {75}},\ \bibinfo {pages}
  {2077} (\bibinfo {year} {1995})}\BibitemShut {NoStop}%
\bibitem [{\citenamefont {Khoury}\ and\ \citenamefont
  {Weltman}(2004)}]{PhysRevLett.93.171104}%
  \BibitemOpen
  \bibfield  {author} {\bibinfo {author} {\bibfnamefont {J.}~\bibnamefont
  {Khoury}}\ and\ \bibinfo {author} {\bibfnamefont {A.}~\bibnamefont
  {Weltman}},\ }\href {\doibase 10.1103/PhysRevLett.93.171104} {\bibfield
  {journal} {\bibinfo  {journal} {Phys. Rev. Lett.}\ }\textbf {\bibinfo
  {volume} {93}},\ \bibinfo {pages} {171104} (\bibinfo {year}
  {2004})}\BibitemShut {NoStop}%
\bibitem [{\citenamefont {Hinterbichler}\ and\ \citenamefont
  {Khoury}(2010)}]{PhysRevLett.104.231301}%
  \BibitemOpen
  \bibfield  {author} {\bibinfo {author} {\bibfnamefont {K.}~\bibnamefont
  {Hinterbichler}}\ and\ \bibinfo {author} {\bibfnamefont {J.}~\bibnamefont
  {Khoury}},\ }\href {\doibase 10.1103/PhysRevLett.104.231301} {\bibfield
  {journal} {\bibinfo  {journal} {Phys. Rev. Lett.}\ }\textbf {\bibinfo
  {volume} {104}},\ \bibinfo {pages} {231301} (\bibinfo {year}
  {2010})}\BibitemShut {NoStop}%
\bibitem [{\citenamefont {Chakrabarti}\ \emph {et~al.}(2022)\citenamefont
  {Chakrabarti}, \citenamefont {Dutta},\ and\ \citenamefont
  {Said}}]{10.1093/mnras/stac1321}%
  \BibitemOpen
  \bibfield  {author} {\bibinfo {author} {\bibfnamefont {S.}~\bibnamefont
  {Chakrabarti}}, \bibinfo {author} {\bibfnamefont {K.}~\bibnamefont {Dutta}},
  \ and\ \bibinfo {author} {\bibfnamefont {J.~L.}\ \bibnamefont {Said}},\
  }\href {\doibase 10.1093/mnras/stac1321} {\bibfield  {journal} {\bibinfo
  {journal} {Monthly Notices of the Royal Astronomical Society}\ }\textbf
  {\bibinfo {volume} {514}},\ \bibinfo {pages} {427} (\bibinfo {year}
  {2022})},\ \Eprint
  {http://arxiv.org/abs/https://academic.oup.com/mnras/article-pdf/514/1/427/43946981/stac1321.pdf}
  {https://academic.oup.com/mnras/article-pdf/514/1/427/43946981/stac1321.pdf}
  \BibitemShut {NoStop}%
\bibitem [{\citenamefont {Dirac}(1937)}]{Dirac}%
  \BibitemOpen
  \bibfield  {author} {\bibinfo {author} {\bibfnamefont {P.~A.~M.}\
  \bibnamefont {Dirac}},\ }\href@noop {} {\bibfield  {journal} {\bibinfo
  {journal} {Nature}\ }\textbf {\bibinfo {volume} {139}},\ \bibinfo {pages}
  {323} (\bibinfo {year} {1937})}\BibitemShut {NoStop}%
\bibitem [{\citenamefont {Jordan}(1937)}]{BD1}%
  \BibitemOpen
  \bibfield  {author} {\bibinfo {author} {\bibfnamefont {P.}~\bibnamefont
  {Jordan}},\ }\href@noop {} {\bibfield  {journal} {\bibinfo  {journal}
  {Naturwissenschaften}\ }\textbf {\bibinfo {volume} {25}},\ \bibinfo {pages}
  {513} (\bibinfo {year} {1937})}\BibitemShut {NoStop}%
\bibitem [{\citenamefont {Fierz}(1956)}]{BD2}%
  \BibitemOpen
  \bibfield  {author} {\bibinfo {author} {\bibfnamefont {M.}~\bibnamefont
  {Fierz}},\ }\href@noop {} {\bibfield  {journal} {\bibinfo  {journal}
  {Helvetica Physica Acta}\ }\textbf {\bibinfo {volume} {29}},\ \bibinfo
  {pages} {128} (\bibinfo {year} {1956})}\BibitemShut {NoStop}%
\bibitem [{\citenamefont {Brans}\ and\ \citenamefont {Dicke}(1961)}]{BD3}%
  \BibitemOpen
  \bibfield  {author} {\bibinfo {author} {\bibfnamefont {C.}~\bibnamefont
  {Brans}}\ and\ \bibinfo {author} {\bibfnamefont {R.~H.}\ \bibnamefont
  {Dicke}},\ }\href {\doibase 10.1103/PhysRev.124.925} {\bibfield  {journal}
  {\bibinfo  {journal} {Phys. Rev.}\ }\textbf {\bibinfo {volume} {124}},\
  \bibinfo {pages} {925} (\bibinfo {year} {1961})}\BibitemShut {NoStop}%
\bibitem [{\citenamefont {Gamow}(1967)}]{gamow}%
  \BibitemOpen
  \bibfield  {author} {\bibinfo {author} {\bibfnamefont {G.}~\bibnamefont
  {Gamow}},\ }\href {\doibase 10.1103/PhysRevLett.19.759} {\bibfield  {journal}
  {\bibinfo  {journal} {Phys. Rev. Lett.}\ }\textbf {\bibinfo {volume} {19}},\
  \bibinfo {pages} {759} (\bibinfo {year} {1967})}\BibitemShut {NoStop}%
\bibitem [{\citenamefont {Bekenstein}(1982)}]{PhysRevD.25.1527}%
  \BibitemOpen
  \bibfield  {author} {\bibinfo {author} {\bibfnamefont {J.~D.}\ \bibnamefont
  {Bekenstein}},\ }\href {\doibase 10.1103/PhysRevD.25.1527} {\bibfield
  {journal} {\bibinfo  {journal} {Phys. Rev. D}\ }\textbf {\bibinfo {volume}
  {25}},\ \bibinfo {pages} {1527} (\bibinfo {year} {1982})}\BibitemShut
  {NoStop}%
\bibitem [{\citenamefont {Uzan}(2003)}]{RevModPhys.75.403}%
  \BibitemOpen
  \bibfield  {author} {\bibinfo {author} {\bibfnamefont {J.-P.}\ \bibnamefont
  {Uzan}},\ }\href {\doibase 10.1103/RevModPhys.75.403} {\bibfield  {journal}
  {\bibinfo  {journal} {Rev. Mod. Phys.}\ }\textbf {\bibinfo {volume} {75}},\
  \bibinfo {pages} {403} (\bibinfo {year} {2003})}\BibitemShut {NoStop}%
\bibitem [{\citenamefont {{Chiba}}(2011)}]{2011PThPh.126..993C}%
  \BibitemOpen
  \bibfield  {author} {\bibinfo {author} {\bibfnamefont {T.}~\bibnamefont
  {{Chiba}}},\ }\href {\doibase 10.1143/PTP.126.993} {\bibfield  {journal}
  {\bibinfo  {journal} {Progress of Theoretical Physics}\ }\textbf {\bibinfo
  {volume} {126}},\ \bibinfo {pages} {993} (\bibinfo {year} {2011})},\ \Eprint
  {http://arxiv.org/abs/1111.0092} {arXiv:1111.0092 [gr-qc]} \BibitemShut
  {NoStop}%
\bibitem [{\citenamefont {Gasser}\ and\ \citenamefont
  {Leutwyler}(1982)}]{GASSER198277}%
  \BibitemOpen
  \bibfield  {author} {\bibinfo {author} {\bibfnamefont {J.}~\bibnamefont
  {Gasser}}\ and\ \bibinfo {author} {\bibfnamefont {H.}~\bibnamefont
  {Leutwyler}},\ }\href {\doibase https://doi.org/10.1016/0370-1573(82)90035-7}
  {\bibfield  {journal} {\bibinfo  {journal} {Physics Reports}\ }\textbf
  {\bibinfo {volume} {87}},\ \bibinfo {pages} {77} (\bibinfo {year}
  {1982})}\BibitemShut {NoStop}%
\bibitem [{\citenamefont {Ji}(1995)}]{PhysRevLett.74.1071}%
  \BibitemOpen
  \bibfield  {author} {\bibinfo {author} {\bibfnamefont {X.}~\bibnamefont
  {Ji}},\ }\href {\doibase 10.1103/PhysRevLett.74.1071} {\bibfield  {journal}
  {\bibinfo  {journal} {Phys. Rev. Lett.}\ }\textbf {\bibinfo {volume} {74}},\
  \bibinfo {pages} {1071} (\bibinfo {year} {1995})}\BibitemShut {NoStop}%
\bibitem [{\citenamefont {Calmet}\ and\ \citenamefont
  {Fritzsch}(2002)}]{Calmet:2002ja}%
  \BibitemOpen
  \bibfield  {author} {\bibinfo {author} {\bibfnamefont {X.}~\bibnamefont
  {Calmet}}\ and\ \bibinfo {author} {\bibfnamefont {H.}~\bibnamefont
  {Fritzsch}},\ }\href {\doibase 10.1016/S0370-2693(02)02147-0} {\bibfield
  {journal} {\bibinfo  {journal} {Phys. Lett. B}\ }\textbf {\bibinfo {volume}
  {540}},\ \bibinfo {pages} {173} (\bibinfo {year} {2002})},\ \Eprint
  {http://arxiv.org/abs/hep-ph/0204258} {arXiv:hep-ph/0204258} \BibitemShut
  {NoStop}%
\bibitem [{\citenamefont {Fritzsch}(2009)}]{FRITZSCH2009221}%
  \BibitemOpen
  \bibfield  {author} {\bibinfo {author} {\bibfnamefont {H.}~\bibnamefont
  {Fritzsch}},\ }\href {\doibase
  https://doi.org/10.1016/j.nuclphysbps.2008.12.050} {\bibfield  {journal}
  {\bibinfo  {journal} {Nuclear Physics B - Proceedings Supplements}\ }\textbf
  {\bibinfo {volume} {186}},\ \bibinfo {pages} {221} (\bibinfo {year}
  {2009})},\ \bibinfo {note} {proceedings of the QCD 08, 14th High-Energy
  Physics International Conference On Quantum ChromoDynamics}\BibitemShut
  {NoStop}%
\bibitem [{\citenamefont {Bagdonaite}\ \emph {et~al.}(2014)\citenamefont
  {Bagdonaite}, \citenamefont {Salumbides}, \citenamefont {Preval},
  \citenamefont {Barstow}, \citenamefont {Barrow}, \citenamefont {Murphy},\
  and\ \citenamefont {Ubachs}}]{PhysRevLett.113.123002}%
  \BibitemOpen
  \bibfield  {author} {\bibinfo {author} {\bibfnamefont {J.}~\bibnamefont
  {Bagdonaite}}, \bibinfo {author} {\bibfnamefont {E.~J.}\ \bibnamefont
  {Salumbides}}, \bibinfo {author} {\bibfnamefont {S.~P.}\ \bibnamefont
  {Preval}}, \bibinfo {author} {\bibfnamefont {M.~A.}\ \bibnamefont {Barstow}},
  \bibinfo {author} {\bibfnamefont {J.~D.}\ \bibnamefont {Barrow}}, \bibinfo
  {author} {\bibfnamefont {M.~T.}\ \bibnamefont {Murphy}}, \ and\ \bibinfo
  {author} {\bibfnamefont {W.}~\bibnamefont {Ubachs}},\ }\href {\doibase
  10.1103/PhysRevLett.113.123002} {\bibfield  {journal} {\bibinfo  {journal}
  {Phys. Rev. Lett.}\ }\textbf {\bibinfo {volume} {113}},\ \bibinfo {pages}
  {123002} (\bibinfo {year} {2014})}\BibitemShut {NoStop}%
\bibitem [{\citenamefont {Huntemann}\ \emph {et~al.}(2014)\citenamefont
  {Huntemann}, \citenamefont {Lipphardt}, \citenamefont {Tamm}, \citenamefont
  {Gerginov}, \citenamefont {Weyers},\ and\ \citenamefont
  {Peik}}]{PhysRevLett.113.210802}%
  \BibitemOpen
  \bibfield  {author} {\bibinfo {author} {\bibfnamefont {N.}~\bibnamefont
  {Huntemann}}, \bibinfo {author} {\bibfnamefont {B.}~\bibnamefont
  {Lipphardt}}, \bibinfo {author} {\bibfnamefont {C.}~\bibnamefont {Tamm}},
  \bibinfo {author} {\bibfnamefont {V.}~\bibnamefont {Gerginov}}, \bibinfo
  {author} {\bibfnamefont {S.}~\bibnamefont {Weyers}}, \ and\ \bibinfo {author}
  {\bibfnamefont {E.}~\bibnamefont {Peik}},\ }\href {\doibase
  10.1103/PhysRevLett.113.210802} {\bibfield  {journal} {\bibinfo  {journal}
  {Phys. Rev. Lett.}\ }\textbf {\bibinfo {volume} {113}},\ \bibinfo {pages}
  {210802} (\bibinfo {year} {2014})}\BibitemShut {NoStop}%
\bibitem [{\citenamefont {Sola}\ \emph {et~al.}(2017)\citenamefont {Sola},
  \citenamefont {Karimkhani},\ and\ \citenamefont
  {Khodam-Mohammadi}}]{Sola:2016our}%
  \BibitemOpen
  \bibfield  {author} {\bibinfo {author} {\bibfnamefont {J.}~\bibnamefont
  {Sola}}, \bibinfo {author} {\bibfnamefont {E.}~\bibnamefont {Karimkhani}}, \
  and\ \bibinfo {author} {\bibfnamefont {A.}~\bibnamefont {Khodam-Mohammadi}},\
  }\href {\doibase 10.1088/1361-6382/34/2/025006} {\bibfield  {journal}
  {\bibinfo  {journal} {Class. Quant. Grav.}\ }\textbf {\bibinfo {volume}
  {34}},\ \bibinfo {pages} {025006} (\bibinfo {year} {2017})},\ \Eprint
  {http://arxiv.org/abs/1609.00350} {arXiv:1609.00350 [gr-qc]} \BibitemShut
  {NoStop}%
\bibitem [{\citenamefont {Chakrabarti}(2021)}]{10.1093/mnras/stab1910}%
  \BibitemOpen
  \bibfield  {author} {\bibinfo {author} {\bibfnamefont {S.}~\bibnamefont
  {Chakrabarti}},\ }\href {\doibase 10.1093/mnras/stab1910} {\bibfield
  {journal} {\bibinfo  {journal} {Monthly Notices of the Royal Astronomical
  Society}\ }\textbf {\bibinfo {volume} {506}},\ \bibinfo {pages} {2518}
  (\bibinfo {year} {2021})},\ \Eprint
  {http://arxiv.org/abs/https://academic.oup.com/mnras/article-pdf/506/2/2518/39136152/stab1910.pdf}
  {https://academic.oup.com/mnras/article-pdf/506/2/2518/39136152/stab1910.pdf}
  \BibitemShut {NoStop}%
\bibitem [{\citenamefont {Bronnikov}\ and\ \citenamefont
  {Fabris}(2006)}]{Bronnikov_2006}%
  \BibitemOpen
  \bibfield  {author} {\bibinfo {author} {\bibfnamefont {K.~A.}\ \bibnamefont
  {Bronnikov}}\ and\ \bibinfo {author} {\bibfnamefont {J.~C.}\ \bibnamefont
  {Fabris}},\ }\href {\doibase 10.1103/physrevlett.96.251101} {\bibfield
  {journal} {\bibinfo  {journal} {Physical Review Letters}\ }\textbf {\bibinfo
  {volume} {96}} (\bibinfo {year} {2006}),\
  10.1103/physrevlett.96.251101}\BibitemShut {NoStop}%
\bibitem [{\citenamefont {Kiselev}(2003)}]{Kiselev_2003}%
  \BibitemOpen
  \bibfield  {author} {\bibinfo {author} {\bibfnamefont {V.~V.}\ \bibnamefont
  {Kiselev}},\ }\href {\doibase 10.1088/0264-9381/20/6/310} {\bibfield
  {journal} {\bibinfo  {journal} {Classical and Quantum Gravity}\ }\textbf
  {\bibinfo {volume} {20}},\ \bibinfo {pages} {1187–1197} (\bibinfo {year}
  {2003})}\BibitemShut {NoStop}%
\bibitem [{\citenamefont {Visser}(2020)}]{Visser_2020}%
  \BibitemOpen
  \bibfield  {author} {\bibinfo {author} {\bibfnamefont {M.}~\bibnamefont
  {Visser}},\ }\href {\doibase 10.1088/1361-6382/ab60b8} {\bibfield  {journal}
  {\bibinfo  {journal} {Classical and Quantum Gravity}\ }\textbf {\bibinfo
  {volume} {37}},\ \bibinfo {pages} {045001} (\bibinfo {year}
  {2020})}\BibitemShut {NoStop}%
\bibitem [{\citenamefont {Battista}\ \emph {et~al.}(2024)\citenamefont
  {Battista}, \citenamefont {Capozziello},\ and\ \citenamefont
  {Errehymy}}]{Battista:2024gud}%
  \BibitemOpen
  \bibfield  {author} {\bibinfo {author} {\bibfnamefont {E.}~\bibnamefont
  {Battista}}, \bibinfo {author} {\bibfnamefont {S.}~\bibnamefont
  {Capozziello}}, \ and\ \bibinfo {author} {\bibfnamefont {A.}~\bibnamefont
  {Errehymy}},\ }\href@noop {} {\  (\bibinfo {year} {2024})},\ \Eprint
  {http://arxiv.org/abs/2409.09750} {arXiv:2409.09750 [gr-qc]} \BibitemShut
  {NoStop}%
\bibitem [{\citenamefont {Di~Grezia}\ \emph {et~al.}(2017)\citenamefont
  {Di~Grezia}, \citenamefont {Battista}, \citenamefont {Manfredonia},\ and\
  \citenamefont {Miele}}]{DiGrezia:2017daq}%
  \BibitemOpen
  \bibfield  {author} {\bibinfo {author} {\bibfnamefont {E.}~\bibnamefont
  {Di~Grezia}}, \bibinfo {author} {\bibfnamefont {E.}~\bibnamefont {Battista}},
  \bibinfo {author} {\bibfnamefont {M.}~\bibnamefont {Manfredonia}}, \ and\
  \bibinfo {author} {\bibfnamefont {G.}~\bibnamefont {Miele}},\ }\href
  {\doibase 10.1140/epjp/i2017-11799-6} {\bibfield  {journal} {\bibinfo
  {journal} {Eur. Phys. J. Plus}\ }\textbf {\bibinfo {volume} {132}},\ \bibinfo
  {pages} {537} (\bibinfo {year} {2017})},\ \Eprint
  {http://arxiv.org/abs/1707.01508} {arXiv:1707.01508 [gr-qc]} \BibitemShut
  {NoStop}%
\bibitem [{\citenamefont {De~Falco}\ \emph {et~al.}(2020)\citenamefont
  {De~Falco}, \citenamefont {Battista}, \citenamefont {Capozziello},\ and\
  \citenamefont {De~Laurentis}}]{DeFalco:2020afv}%
  \BibitemOpen
  \bibfield  {author} {\bibinfo {author} {\bibfnamefont {V.}~\bibnamefont
  {De~Falco}}, \bibinfo {author} {\bibfnamefont {E.}~\bibnamefont {Battista}},
  \bibinfo {author} {\bibfnamefont {S.}~\bibnamefont {Capozziello}}, \ and\
  \bibinfo {author} {\bibfnamefont {M.}~\bibnamefont {De~Laurentis}},\ }\href
  {\doibase 10.1103/PhysRevD.101.104037} {\bibfield  {journal} {\bibinfo
  {journal} {Phys. Rev. D}\ }\textbf {\bibinfo {volume} {101}},\ \bibinfo
  {pages} {104037} (\bibinfo {year} {2020})},\ \Eprint
  {http://arxiv.org/abs/2004.14849} {arXiv:2004.14849 [gr-qc]} \BibitemShut
  {NoStop}%
\bibitem [{\citenamefont {De~Falco}\ \emph
  {et~al.}(2021{\natexlab{a}})\citenamefont {De~Falco}, \citenamefont
  {Battista}, \citenamefont {Capozziello},\ and\ \citenamefont
  {De~Laurentis}}]{DeFalco:2021klh}%
  \BibitemOpen
  \bibfield  {author} {\bibinfo {author} {\bibfnamefont {V.}~\bibnamefont
  {De~Falco}}, \bibinfo {author} {\bibfnamefont {E.}~\bibnamefont {Battista}},
  \bibinfo {author} {\bibfnamefont {S.}~\bibnamefont {Capozziello}}, \ and\
  \bibinfo {author} {\bibfnamefont {M.}~\bibnamefont {De~Laurentis}},\ }\href
  {\doibase 10.1103/PhysRevD.103.044007} {\bibfield  {journal} {\bibinfo
  {journal} {Phys. Rev. D}\ }\textbf {\bibinfo {volume} {103}},\ \bibinfo
  {pages} {044007} (\bibinfo {year} {2021}{\natexlab{a}})},\ \Eprint
  {http://arxiv.org/abs/2101.04960} {arXiv:2101.04960 [gr-qc]} \BibitemShut
  {NoStop}%
\bibitem [{\citenamefont {De~Falco}\ \emph
  {et~al.}(2021{\natexlab{b}})\citenamefont {De~Falco}, \citenamefont
  {Battista}, \citenamefont {Capozziello},\ and\ \citenamefont
  {De~Laurentis}}]{DeFalco:2021ksd}%
  \BibitemOpen
  \bibfield  {author} {\bibinfo {author} {\bibfnamefont {V.}~\bibnamefont
  {De~Falco}}, \bibinfo {author} {\bibfnamefont {E.}~\bibnamefont {Battista}},
  \bibinfo {author} {\bibfnamefont {S.}~\bibnamefont {Capozziello}}, \ and\
  \bibinfo {author} {\bibfnamefont {M.}~\bibnamefont {De~Laurentis}},\ }\href
  {\doibase 10.1140/epjc/s10052-021-08958-4} {\bibfield  {journal} {\bibinfo
  {journal} {Eur. Phys. J. C}\ }\textbf {\bibinfo {volume} {81}},\ \bibinfo
  {pages} {157} (\bibinfo {year} {2021}{\natexlab{b}})},\ \Eprint
  {http://arxiv.org/abs/2102.01123} {arXiv:2102.01123 [gr-qc]} \BibitemShut
  {NoStop}%
\bibitem [{\citenamefont {Chew}\ and\ \citenamefont
  {Yeom}(2024)}]{Chew:2024rin}%
  \BibitemOpen
  \bibfield  {author} {\bibinfo {author} {\bibfnamefont {X.~Y.}\ \bibnamefont
  {Chew}}\ and\ \bibinfo {author} {\bibfnamefont {D.-h.}\ \bibnamefont
  {Yeom}},\ }\href {\doibase 10.1103/PhysRevD.110.044036} {\bibfield  {journal}
  {\bibinfo  {journal} {Phys. Rev. D}\ }\textbf {\bibinfo {volume} {110}},\
  \bibinfo {pages} {044036} (\bibinfo {year} {2024})},\ \Eprint
  {http://arxiv.org/abs/2401.09039} {arXiv:2401.09039 [gr-qc]} \BibitemShut
  {NoStop}%
\bibitem [{\citenamefont {Chew}\ \emph {et~al.}(2023)\citenamefont {Chew},
  \citenamefont {Yeom},\ and\ \citenamefont
  {Bl\'azquez-Salcedo}}]{Chew:2022enh}%
  \BibitemOpen
  \bibfield  {author} {\bibinfo {author} {\bibfnamefont {X.~Y.}\ \bibnamefont
  {Chew}}, \bibinfo {author} {\bibfnamefont {D.-h.}\ \bibnamefont {Yeom}}, \
  and\ \bibinfo {author} {\bibfnamefont {J.~L.}\ \bibnamefont
  {Bl\'azquez-Salcedo}},\ }\href {\doibase 10.1103/PhysRevD.108.044020}
  {\bibfield  {journal} {\bibinfo  {journal} {Phys. Rev. D}\ }\textbf {\bibinfo
  {volume} {108}},\ \bibinfo {pages} {044020} (\bibinfo {year} {2023})},\
  \Eprint {http://arxiv.org/abs/2210.01313} {arXiv:2210.01313 [gr-qc]}
  \BibitemShut {NoStop}%
\bibitem [{\citenamefont {Chew}\ and\ \citenamefont
  {Lim}(2024{\natexlab{a}})}]{Chew:2023olq}%
  \BibitemOpen
  \bibfield  {author} {\bibinfo {author} {\bibfnamefont {X.~Y.}\ \bibnamefont
  {Chew}}\ and\ \bibinfo {author} {\bibfnamefont {K.-G.}\ \bibnamefont {Lim}},\
  }\href {\doibase 10.1103/PhysRevD.109.064039} {\bibfield  {journal} {\bibinfo
   {journal} {Phys. Rev. D}\ }\textbf {\bibinfo {volume} {109}},\ \bibinfo
  {pages} {064039} (\bibinfo {year} {2024}{\natexlab{a}})},\ \Eprint
  {http://arxiv.org/abs/2307.13972} {arXiv:2307.13972 [gr-qc]} \BibitemShut
  {NoStop}%
\bibitem [{\citenamefont {Chew}\ and\ \citenamefont
  {Lim}(2024{\natexlab{b}})}]{Chew:2024bec}%
  \BibitemOpen
  \bibfield  {author} {\bibinfo {author} {\bibfnamefont {X.~Y.}\ \bibnamefont
  {Chew}}\ and\ \bibinfo {author} {\bibfnamefont {K.-G.}\ \bibnamefont {Lim}},\
  }\href {\doibase 10.3390/universe10050212} {\bibfield  {journal} {\bibinfo
  {journal} {Universe}\ }\textbf {\bibinfo {volume} {10}},\ \bibinfo {pages}
  {212} (\bibinfo {year} {2024}{\natexlab{b}})},\ \Eprint
  {http://arxiv.org/abs/2405.06407} {arXiv:2405.06407 [gr-qc]} \BibitemShut
  {NoStop}%
\bibitem [{\citenamefont {Chew}\ and\ \citenamefont
  {Myung}(2024)}]{Chew:2024evh}%
  \BibitemOpen
  \bibfield  {author} {\bibinfo {author} {\bibfnamefont {X.~Y.}\ \bibnamefont
  {Chew}}\ and\ \bibinfo {author} {\bibfnamefont {Y.~S.}\ \bibnamefont
  {Myung}},\ }\href {\doibase 10.1103/PhysRevD.110.044011} {\bibfield
  {journal} {\bibinfo  {journal} {Phys. Rev. D}\ }\textbf {\bibinfo {volume}
  {110}},\ \bibinfo {pages} {044011} (\bibinfo {year} {2024})},\ \Eprint
  {http://arxiv.org/abs/2405.04921} {arXiv:2405.04921 [gr-qc]} \BibitemShut
  {NoStop}%
\bibitem [{\citenamefont {Dzhunushaliev}\ \emph {et~al.}(2008)\citenamefont
  {Dzhunushaliev}, \citenamefont {Folomeev}, \citenamefont {Myrzakulov},\ and\
  \citenamefont {Singleton}}]{Dzhunushaliev:2008bq}%
  \BibitemOpen
  \bibfield  {author} {\bibinfo {author} {\bibfnamefont {V.}~\bibnamefont
  {Dzhunushaliev}}, \bibinfo {author} {\bibfnamefont {V.}~\bibnamefont
  {Folomeev}}, \bibinfo {author} {\bibfnamefont {R.}~\bibnamefont
  {Myrzakulov}}, \ and\ \bibinfo {author} {\bibfnamefont {D.}~\bibnamefont
  {Singleton}},\ }\href {\doibase 10.1088/1126-6708/2008/07/094} {\bibfield
  {journal} {\bibinfo  {journal} {JHEP}\ }\textbf {\bibinfo {volume} {07}},\
  \bibinfo {pages} {094} (\bibinfo {year} {2008})},\ \Eprint
  {http://arxiv.org/abs/0805.3211} {arXiv:0805.3211 [gr-qc]} \BibitemShut
  {NoStop}%
\bibitem [{\citenamefont {Pereira}\ \emph
  {et~al.}(2024{\natexlab{a}})\citenamefont {Pereira}, \citenamefont
  {Rodrigues}, \citenamefont {Fabris},\ and\ \citenamefont
  {Rodrigues}}]{PhysRevD.109.044011}%
  \BibitemOpen
  \bibfield  {author} {\bibinfo {author} {\bibfnamefont {C.~F.~S.}\
  \bibnamefont {Pereira}}, \bibinfo {author} {\bibfnamefont {D.~C.}\
  \bibnamefont {Rodrigues}}, \bibinfo {author} {\bibfnamefont {J.~C.}\
  \bibnamefont {Fabris}}, \ and\ \bibinfo {author} {\bibfnamefont {M.~E.}\
  \bibnamefont {Rodrigues}},\ }\href {\doibase 10.1103/PhysRevD.109.044011}
  {\bibfield  {journal} {\bibinfo  {journal} {Phys. Rev. D}\ }\textbf {\bibinfo
  {volume} {109}},\ \bibinfo {pages} {044011} (\bibinfo {year}
  {2024}{\natexlab{a}})}\BibitemShut {NoStop}%
\bibitem [{\citenamefont {Pereira}\ \emph
  {et~al.}(2024{\natexlab{b}})\citenamefont {Pereira}, \citenamefont {Martins},
  \citenamefont {C.~Rodrigues}, \citenamefont {Fabris},\ and\ \citenamefont
  {Rodrigues}}]{Pereira:2024gsl}%
  \BibitemOpen
  \bibfield  {author} {\bibinfo {author} {\bibfnamefont {C.~F.~S.}\
  \bibnamefont {Pereira}}, \bibinfo {author} {\bibfnamefont {E.~L.}\
  \bibnamefont {Martins}}, \bibinfo {author} {\bibfnamefont {D.}~\bibnamefont
  {C.~Rodrigues}}, \bibinfo {author} {\bibfnamefont {J.~C.}\ \bibnamefont
  {Fabris}}, \ and\ \bibinfo {author} {\bibfnamefont {M.~E.}\ \bibnamefont
  {Rodrigues}},\ }\href@noop {} {\  (\bibinfo {year} {2024}{\natexlab{b}})},\
  \Eprint {http://arxiv.org/abs/2405.07455} {arXiv:2405.07455 [gr-qc]}
  \BibitemShut {NoStop}%
\bibitem [{\citenamefont {Pereira}\ \emph
  {et~al.}(2024{\natexlab{c}})\citenamefont {Pereira}, \citenamefont
  {C.~Rodrigues}, \citenamefont {Silva}, \citenamefont {Fabris}, \citenamefont
  {Rodrigues},\ and\ \citenamefont {Belich}}]{Pereira:2024rtv}%
  \BibitemOpen
  \bibfield  {author} {\bibinfo {author} {\bibfnamefont {C.~F.~S.}\
  \bibnamefont {Pereira}}, \bibinfo {author} {\bibfnamefont {D.}~\bibnamefont
  {C.~Rodrigues}}, \bibinfo {author} {\bibfnamefont {M.~V. d.~S.}\ \bibnamefont
  {Silva}}, \bibinfo {author} {\bibfnamefont {J.~C.}\ \bibnamefont {Fabris}},
  \bibinfo {author} {\bibfnamefont {M.~E.}\ \bibnamefont {Rodrigues}}, \ and\
  \bibinfo {author} {\bibfnamefont {H.}~\bibnamefont {Belich}},\ }\href@noop {}
  {\  (\bibinfo {year} {2024}{\natexlab{c}})},\ \Eprint
  {http://arxiv.org/abs/2409.09182} {arXiv:2409.09182 [gr-qc]} \BibitemShut
  {NoStop}%
\bibitem [{\citenamefont {Bronnikov}\ \emph {et~al.}(2021)\citenamefont
  {Bronnikov}, \citenamefont {Konoplya},\ and\ \citenamefont
  {Pappas}}]{Bronnikov:2021liv}%
  \BibitemOpen
  \bibfield  {author} {\bibinfo {author} {\bibfnamefont {K.~A.}\ \bibnamefont
  {Bronnikov}}, \bibinfo {author} {\bibfnamefont {R.~A.}\ \bibnamefont
  {Konoplya}}, \ and\ \bibinfo {author} {\bibfnamefont {T.~D.}\ \bibnamefont
  {Pappas}},\ }\href {\doibase 10.1103/PhysRevD.103.124062} {\bibfield
  {journal} {\bibinfo  {journal} {Phys. Rev. D}\ }\textbf {\bibinfo {volume}
  {103}},\ \bibinfo {pages} {124062} (\bibinfo {year} {2021})},\ \Eprint
  {http://arxiv.org/abs/2102.10679} {arXiv:2102.10679 [gr-qc]} \BibitemShut
  {NoStop}%
\bibitem [{\citenamefont {Simpson}\ and\ \citenamefont
  {Visser}(2019)}]{Simpson_2019}%
  \BibitemOpen
  \bibfield  {author} {\bibinfo {author} {\bibfnamefont {A.}~\bibnamefont
  {Simpson}}\ and\ \bibinfo {author} {\bibfnamefont {M.}~\bibnamefont
  {Visser}},\ }\href {\doibase 10.1088/1475-7516/2019/02/042} {\bibfield
  {journal} {\bibinfo  {journal} {Journal of Cosmology and Astroparticle
  Physics}\ }\textbf {\bibinfo {volume} {2019}},\ \bibinfo {pages} {042}
  (\bibinfo {year} {2019})}\BibitemShut {NoStop}%
\bibitem [{\citenamefont {Chakrabarti}\ and\ \citenamefont
  {Kar}(2021)}]{PhysRevD.104.024071}%
  \BibitemOpen
  \bibfield  {author} {\bibinfo {author} {\bibfnamefont {S.}~\bibnamefont
  {Chakrabarti}}\ and\ \bibinfo {author} {\bibfnamefont {S.}~\bibnamefont
  {Kar}},\ }\href {\doibase 10.1103/PhysRevD.104.024071} {\bibfield  {journal}
  {\bibinfo  {journal} {Phys. Rev. D}\ }\textbf {\bibinfo {volume} {104}},\
  \bibinfo {pages} {024071} (\bibinfo {year} {2021})}\BibitemShut {NoStop}%
\bibitem [{\citenamefont {Visser}(1995)}]{Visser:1995cc}%
  \BibitemOpen
  \bibfield  {author} {\bibinfo {author} {\bibfnamefont {M.}~\bibnamefont
  {Visser}},\ }\href@noop {} {\emph {\bibinfo {title} {{Lorentzian wormholes:
  From Einstein to Hawking}}}}\ (\bibinfo {year} {1995})\BibitemShut {NoStop}%
\bibitem [{\citenamefont {Einstein}\ and\ \citenamefont
  {Rosen}(1935)}]{PhysRev.48.73}%
  \BibitemOpen
  \bibfield  {author} {\bibinfo {author} {\bibfnamefont {A.}~\bibnamefont
  {Einstein}}\ and\ \bibinfo {author} {\bibfnamefont {N.}~\bibnamefont
  {Rosen}},\ }\href {\doibase 10.1103/PhysRev.48.73} {\bibfield  {journal}
  {\bibinfo  {journal} {Phys. Rev.}\ }\textbf {\bibinfo {volume} {48}},\
  \bibinfo {pages} {73} (\bibinfo {year} {1935})}\BibitemShut {NoStop}%
\bibitem [{\citenamefont {Misner}\ and\ \citenamefont
  {Wheeler}(1957)}]{MISNER1957525}%
  \BibitemOpen
  \bibfield  {author} {\bibinfo {author} {\bibfnamefont {C.~W.}\ \bibnamefont
  {Misner}}\ and\ \bibinfo {author} {\bibfnamefont {J.~A.}\ \bibnamefont
  {Wheeler}},\ }\href {\doibase https://doi.org/10.1016/0003-4916(57)90049-0}
  {\bibfield  {journal} {\bibinfo  {journal} {Annals of Physics}\ }\textbf
  {\bibinfo {volume} {2}},\ \bibinfo {pages} {525} (\bibinfo {year}
  {1957})}\BibitemShut {NoStop}%
\bibitem [{\citenamefont {Kolassis}\ \emph {et~al.}(1988)\citenamefont
  {Kolassis}, \citenamefont {Santos},\ and\ \citenamefont
  {Tsoubelis}}]{Kolassis_1988}%
  \BibitemOpen
  \bibfield  {author} {\bibinfo {author} {\bibfnamefont {C.~A.}\ \bibnamefont
  {Kolassis}}, \bibinfo {author} {\bibfnamefont {N.~O.}\ \bibnamefont
  {Santos}}, \ and\ \bibinfo {author} {\bibfnamefont {D.}~\bibnamefont
  {Tsoubelis}},\ }\href {\doibase 10.1088/0264-9381/5/10/011} {\bibfield
  {journal} {\bibinfo  {journal} {Classical and Quantum Gravity}\ }\textbf
  {\bibinfo {volume} {5}},\ \bibinfo {pages} {1329} (\bibinfo {year}
  {1988})}\BibitemShut {NoStop}%
\bibitem [{\citenamefont {Ellis}(1973)}]{10.10631.1666161}%
  \BibitemOpen
  \bibfield  {author} {\bibinfo {author} {\bibfnamefont {H.~G.}\ \bibnamefont
  {Ellis}},\ }\href {\doibase 10.1063/1.1666161} {\bibfield  {journal}
  {\bibinfo  {journal} {Journal of Mathematical Physics}\ }\textbf {\bibinfo
  {volume} {14}},\ \bibinfo {pages} {104} (\bibinfo {year} {1973})},\ \Eprint
  {http://arxiv.org/abs/https://pubs.aip.org/aip/jmp/article-pdf/14/1/104/19133700/104\_1\_online.pdf}
  {https://pubs.aip.org/aip/jmp/article-pdf/14/1/104/19133700/104\_1\_online.pdf}
  \BibitemShut {NoStop}%
\bibitem [{\citenamefont {Brax}\ \emph {et~al.}(2010)\citenamefont {Brax},
  \citenamefont {van~de Bruck}, \citenamefont {Mota}, \citenamefont {Nunes},\
  and\ \citenamefont {Winther}}]{PhysRevD.82.083503}%
  \BibitemOpen
  \bibfield  {author} {\bibinfo {author} {\bibfnamefont {P.}~\bibnamefont
  {Brax}}, \bibinfo {author} {\bibfnamefont {C.}~\bibnamefont {van~de Bruck}},
  \bibinfo {author} {\bibfnamefont {D.~F.}\ \bibnamefont {Mota}}, \bibinfo
  {author} {\bibfnamefont {N.~J.}\ \bibnamefont {Nunes}}, \ and\ \bibinfo
  {author} {\bibfnamefont {H.~A.}\ \bibnamefont {Winther}},\ }\href {\doibase
  10.1103/PhysRevD.82.083503} {\bibfield  {journal} {\bibinfo  {journal} {Phys.
  Rev. D}\ }\textbf {\bibinfo {volume} {82}},\ \bibinfo {pages} {083503}
  (\bibinfo {year} {2010})}\BibitemShut {NoStop}%
\bibitem [{\citenamefont {Hinterbichler}\ \emph {et~al.}(2011)\citenamefont
  {Hinterbichler}, \citenamefont {Khoury}, \citenamefont {Levy},\ and\
  \citenamefont {Matas}}]{PhysRevD.84.103521}%
  \BibitemOpen
  \bibfield  {author} {\bibinfo {author} {\bibfnamefont {K.}~\bibnamefont
  {Hinterbichler}}, \bibinfo {author} {\bibfnamefont {J.}~\bibnamefont
  {Khoury}}, \bibinfo {author} {\bibfnamefont {A.}~\bibnamefont {Levy}}, \ and\
  \bibinfo {author} {\bibfnamefont {A.}~\bibnamefont {Matas}},\ }\href
  {\doibase 10.1103/PhysRevD.84.103521} {\bibfield  {journal} {\bibinfo
  {journal} {Phys. Rev. D}\ }\textbf {\bibinfo {volume} {84}},\ \bibinfo
  {pages} {103521} (\bibinfo {year} {2011})}\BibitemShut {NoStop}%
\bibitem [{\citenamefont {Virbhadra}\ and\ \citenamefont
  {Ellis}(2000)}]{Virbhadra:1999nm}%
  \BibitemOpen
  \bibfield  {author} {\bibinfo {author} {\bibfnamefont {K.~S.}\ \bibnamefont
  {Virbhadra}}\ and\ \bibinfo {author} {\bibfnamefont {G.~F.~R.}\ \bibnamefont
  {Ellis}},\ }\href {\doibase 10.1103/PhysRevD.62.084003} {\bibfield  {journal}
  {\bibinfo  {journal} {Phys. Rev. D}\ }\textbf {\bibinfo {volume} {62}},\
  \bibinfo {pages} {084003} (\bibinfo {year} {2000})},\ \Eprint
  {http://arxiv.org/abs/astro-ph/9904193} {arXiv:astro-ph/9904193} \BibitemShut
  {NoStop}%
\bibitem [{\citenamefont {Virbhadra}\ and\ \citenamefont
  {Ellis}(2002)}]{PhysRevD.65.103004}%
  \BibitemOpen
  \bibfield  {author} {\bibinfo {author} {\bibfnamefont {K.~S.}\ \bibnamefont
  {Virbhadra}}\ and\ \bibinfo {author} {\bibfnamefont {G.~F.~R.}\ \bibnamefont
  {Ellis}},\ }\href {\doibase 10.1103/PhysRevD.65.103004} {\bibfield  {journal}
  {\bibinfo  {journal} {Phys. Rev. D}\ }\textbf {\bibinfo {volume} {65}},\
  \bibinfo {pages} {103004} (\bibinfo {year} {2002})}\BibitemShut {NoStop}%
\bibitem [{\citenamefont {Virbhadra}\ and\ \citenamefont
  {Keeton}(2008)}]{PhysRevD.77.124014}%
  \BibitemOpen
  \bibfield  {author} {\bibinfo {author} {\bibfnamefont {K.~S.}\ \bibnamefont
  {Virbhadra}}\ and\ \bibinfo {author} {\bibfnamefont {C.~R.}\ \bibnamefont
  {Keeton}},\ }\href {\doibase 10.1103/PhysRevD.77.124014} {\bibfield
  {journal} {\bibinfo  {journal} {Phys. Rev. D}\ }\textbf {\bibinfo {volume}
  {77}},\ \bibinfo {pages} {124014} (\bibinfo {year} {2008})}\BibitemShut
  {NoStop}%
\bibitem [{\citenamefont {Claudel}\ \emph {et~al.}(2001)\citenamefont
  {Claudel}, \citenamefont {Virbhadra},\ and\ \citenamefont
  {Ellis}}]{10.1063/1.1308507}%
  \BibitemOpen
  \bibfield  {author} {\bibinfo {author} {\bibfnamefont {C.-M.}\ \bibnamefont
  {Claudel}}, \bibinfo {author} {\bibfnamefont {K.~S.}\ \bibnamefont
  {Virbhadra}}, \ and\ \bibinfo {author} {\bibfnamefont {G.~F.~R.}\
  \bibnamefont {Ellis}},\ }\href {\doibase 10.1063/1.1308507} {\bibfield
  {journal} {\bibinfo  {journal} {Journal of Mathematical Physics}\ }\textbf
  {\bibinfo {volume} {42}},\ \bibinfo {pages} {818} (\bibinfo {year} {2001})},\
  \Eprint
  {http://arxiv.org/abs/https://pubs.aip.org/aip/jmp/article-pdf/42/2/818/19220437/818\_1\_online.pdf}
  {https://pubs.aip.org/aip/jmp/article-pdf/42/2/818/19220437/818\_1\_online.pdf}
  \BibitemShut {NoStop}%
\bibitem [{\citenamefont {Virbhadra}(2009)}]{PhysRevD.79.083004}%
  \BibitemOpen
  \bibfield  {author} {\bibinfo {author} {\bibfnamefont {K.~S.}\ \bibnamefont
  {Virbhadra}},\ }\href {\doibase 10.1103/PhysRevD.79.083004} {\bibfield
  {journal} {\bibinfo  {journal} {Phys. Rev. D}\ }\textbf {\bibinfo {volume}
  {79}},\ \bibinfo {pages} {083004} (\bibinfo {year} {2009})}\BibitemShut
  {NoStop}%
\bibitem [{\citenamefont {Wald}(1984)}]{Wald:1984rg}%
  \BibitemOpen
  \bibfield  {author} {\bibinfo {author} {\bibfnamefont {R.~M.}\ \bibnamefont
  {Wald}},\ }\href {\doibase 10.7208/chicago/9780226870373.001.0001} {\emph
  {\bibinfo {title} {{General Relativity}}}}\ (\bibinfo  {publisher} {Chicago
  Univ. Pr.},\ \bibinfo {address} {Chicago, USA},\ \bibinfo {year}
  {1984})\BibitemShut {NoStop}%
\bibitem [{\citenamefont {Chakraborty}(2021)}]{Chakraborty:2021dmu}%
  \BibitemOpen
  \bibfield  {author} {\bibinfo {author} {\bibfnamefont {S.}~\bibnamefont
  {Chakraborty}},\ }\href {\doibase 10.3390/galaxies9040096} {\bibfield
  {journal} {\bibinfo  {journal} {Galaxies}\ }\textbf {\bibinfo {volume} {9}},\
  \bibinfo {pages} {96} (\bibinfo {year} {2021})},\ \Eprint
  {http://arxiv.org/abs/2111.04912} {arXiv:2111.04912 [gr-qc]} \BibitemShut
  {NoStop}%
\bibitem [{\citenamefont {Mishra}\ \emph {et~al.}(2019)\citenamefont {Mishra},
  \citenamefont {Chakraborty},\ and\ \citenamefont {Sarkar}}]{Mishra:2019trb}%
  \BibitemOpen
  \bibfield  {author} {\bibinfo {author} {\bibfnamefont {A.~K.}\ \bibnamefont
  {Mishra}}, \bibinfo {author} {\bibfnamefont {S.}~\bibnamefont {Chakraborty}},
  \ and\ \bibinfo {author} {\bibfnamefont {S.}~\bibnamefont {Sarkar}},\ }\href
  {\doibase 10.1103/PhysRevD.99.104080} {\bibfield  {journal} {\bibinfo
  {journal} {Phys. Rev. D}\ }\textbf {\bibinfo {volume} {99}},\ \bibinfo
  {pages} {104080} (\bibinfo {year} {2019})},\ \Eprint
  {http://arxiv.org/abs/1903.06376} {arXiv:1903.06376 [gr-qc]} \BibitemShut
  {NoStop}%
\bibitem [{\citenamefont {Berry}\ \emph {et~al.}(2020)\citenamefont {Berry},
  \citenamefont {Simpson},\ and\ \citenamefont {Visser}}]{Berry:2020ntz}%
  \BibitemOpen
  \bibfield  {author} {\bibinfo {author} {\bibfnamefont {T.}~\bibnamefont
  {Berry}}, \bibinfo {author} {\bibfnamefont {A.}~\bibnamefont {Simpson}}, \
  and\ \bibinfo {author} {\bibfnamefont {M.}~\bibnamefont {Visser}},\ }\href
  {\doibase 10.3390/universe7010002} {\bibfield  {journal} {\bibinfo  {journal}
  {Universe}\ }\textbf {\bibinfo {volume} {7}},\ \bibinfo {pages} {2} (\bibinfo
  {year} {2020})},\ \Eprint {http://arxiv.org/abs/2008.13308} {arXiv:2008.13308
  [gr-qc]} \BibitemShut {NoStop}%
\bibitem [{\citenamefont {Tang}\ \emph {et~al.}(2017)\citenamefont {Tang},
  \citenamefont {Ong},\ and\ \citenamefont {Wang}}]{Tang:2017enb}%
  \BibitemOpen
  \bibfield  {author} {\bibinfo {author} {\bibfnamefont {Z.-Y.}\ \bibnamefont
  {Tang}}, \bibinfo {author} {\bibfnamefont {Y.~C.}\ \bibnamefont {Ong}}, \
  and\ \bibinfo {author} {\bibfnamefont {B.}~\bibnamefont {Wang}},\ }\href
  {\doibase 10.1088/1361-6382/aa95ff} {\bibfield  {journal} {\bibinfo
  {journal} {Class. Quant. Grav.}\ }\textbf {\bibinfo {volume} {34}},\ \bibinfo
  {pages} {245006} (\bibinfo {year} {2017})},\ \Eprint
  {http://arxiv.org/abs/1705.09633} {arXiv:1705.09633 [gr-qc]} \BibitemShut
  {NoStop}%
\bibitem [{\citenamefont {Mishra}\ and\ \citenamefont
  {Chakraborty}(2020)}]{Mishra:2020jlw}%
  \BibitemOpen
  \bibfield  {author} {\bibinfo {author} {\bibfnamefont {A.~K.}\ \bibnamefont
  {Mishra}}\ and\ \bibinfo {author} {\bibfnamefont {S.}~\bibnamefont
  {Chakraborty}},\ }\href {\doibase 10.1103/PhysRevD.101.064041} {\bibfield
  {journal} {\bibinfo  {journal} {Phys. Rev. D}\ }\textbf {\bibinfo {volume}
  {101}},\ \bibinfo {pages} {064041} (\bibinfo {year} {2020})},\ \Eprint
  {http://arxiv.org/abs/1911.09855} {arXiv:1911.09855 [gr-qc]} \BibitemShut
  {NoStop}%
\bibitem [{\citenamefont {Rahman}\ \emph {et~al.}(2019)\citenamefont {Rahman},
  \citenamefont {Chakraborty}, \citenamefont {SenGupta},\ and\ \citenamefont
  {Sen}}]{Rahman:2018oso}%
  \BibitemOpen
  \bibfield  {author} {\bibinfo {author} {\bibfnamefont {M.}~\bibnamefont
  {Rahman}}, \bibinfo {author} {\bibfnamefont {S.}~\bibnamefont {Chakraborty}},
  \bibinfo {author} {\bibfnamefont {S.}~\bibnamefont {SenGupta}}, \ and\
  \bibinfo {author} {\bibfnamefont {A.~A.}\ \bibnamefont {Sen}},\ }\href
  {\doibase 10.1007/JHEP03(2019)178} {\bibfield  {journal} {\bibinfo  {journal}
  {JHEP}\ }\textbf {\bibinfo {volume} {03}},\ \bibinfo {pages} {178} (\bibinfo
  {year} {2019})},\ \Eprint {http://arxiv.org/abs/1811.08538} {arXiv:1811.08538
  [gr-qc]} \BibitemShut {NoStop}%
\bibitem [{\citenamefont {Cardoso}\ \emph {et~al.}(2018)\citenamefont
  {Cardoso}, \citenamefont {Costa}, \citenamefont {Destounis}, \citenamefont
  {Hintz},\ and\ \citenamefont {Jansen}}]{Cardoso:2017soq}%
  \BibitemOpen
  \bibfield  {author} {\bibinfo {author} {\bibfnamefont {V.}~\bibnamefont
  {Cardoso}}, \bibinfo {author} {\bibfnamefont {J.~a.~L.}\ \bibnamefont
  {Costa}}, \bibinfo {author} {\bibfnamefont {K.}~\bibnamefont {Destounis}},
  \bibinfo {author} {\bibfnamefont {P.}~\bibnamefont {Hintz}}, \ and\ \bibinfo
  {author} {\bibfnamefont {A.}~\bibnamefont {Jansen}},\ }\href {\doibase
  10.1103/PhysRevLett.120.031103} {\bibfield  {journal} {\bibinfo  {journal}
  {Phys. Rev. Lett.}\ }\textbf {\bibinfo {volume} {120}},\ \bibinfo {pages}
  {031103} (\bibinfo {year} {2018})},\ \Eprint
  {http://arxiv.org/abs/1711.10502} {arXiv:1711.10502 [gr-qc]} \BibitemShut
  {NoStop}%
\bibitem [{\citenamefont {Cardoso}\ \emph {et~al.}(2009)\citenamefont
  {Cardoso}, \citenamefont {Miranda}, \citenamefont {Berti}, \citenamefont
  {Witek},\ and\ \citenamefont {Zanchin}}]{Cardoso:2008bp}%
  \BibitemOpen
  \bibfield  {author} {\bibinfo {author} {\bibfnamefont {V.}~\bibnamefont
  {Cardoso}}, \bibinfo {author} {\bibfnamefont {A.~S.}\ \bibnamefont
  {Miranda}}, \bibinfo {author} {\bibfnamefont {E.}~\bibnamefont {Berti}},
  \bibinfo {author} {\bibfnamefont {H.}~\bibnamefont {Witek}}, \ and\ \bibinfo
  {author} {\bibfnamefont {V.~T.}\ \bibnamefont {Zanchin}},\ }\href {\doibase
  10.1103/PhysRevD.79.064016} {\bibfield  {journal} {\bibinfo  {journal} {Phys.
  Rev. D}\ }\textbf {\bibinfo {volume} {79}},\ \bibinfo {pages} {064016}
  (\bibinfo {year} {2009})},\ \Eprint {http://arxiv.org/abs/0812.1806}
  {arXiv:0812.1806 [hep-th]} \BibitemShut {NoStop}%
\bibitem [{\citenamefont {Chandrasekhar}(1983)}]{book:338838}%
  \BibitemOpen
  \bibfield  {author} {\bibinfo {author} {\bibfnamefont {S.}~\bibnamefont
  {Chandrasekhar}},\ }\href
  {http://gen.lib.rus.ec/book/index.php?md5=412af084a55f8e3768e551eb3505db2a}
  {\emph {\bibinfo {title} {The Mathematical Theory of Black Holes}}},\ The
  International Series of Monographs on Physics\ (\bibinfo  {publisher} {Oxford
  University Press},\ \bibinfo {year} {1983})\BibitemShut {NoStop}%
\bibitem [{\citenamefont {Visser}(1992)}]{PhysRevD.46.2445}%
  \BibitemOpen
  \bibfield  {author} {\bibinfo {author} {\bibfnamefont {M.}~\bibnamefont
  {Visser}},\ }\href {\doibase 10.1103/PhysRevD.46.2445} {\bibfield  {journal}
  {\bibinfo  {journal} {Phys. Rev. D}\ }\textbf {\bibinfo {volume} {46}},\
  \bibinfo {pages} {2445} (\bibinfo {year} {1992})}\BibitemShut {NoStop}%
\bibitem [{\citenamefont {Visser}(1993)}]{PhysRevD.48.583}%
  \BibitemOpen
  \bibfield  {author} {\bibinfo {author} {\bibfnamefont {M.}~\bibnamefont
  {Visser}},\ }\href {\doibase 10.1103/PhysRevD.48.583} {\bibfield  {journal}
  {\bibinfo  {journal} {Phys. Rev. D}\ }\textbf {\bibinfo {volume} {48}},\
  \bibinfo {pages} {583} (\bibinfo {year} {1993})}\BibitemShut {NoStop}%
\bibitem [{\citenamefont {Boonserm}\ \emph {et~al.}(2013)\citenamefont
  {Boonserm}, \citenamefont {Ngampitipan},\ and\ \citenamefont
  {Visser}}]{PhysRevD.88.041502}%
  \BibitemOpen
  \bibfield  {author} {\bibinfo {author} {\bibfnamefont {P.}~\bibnamefont
  {Boonserm}}, \bibinfo {author} {\bibfnamefont {T.}~\bibnamefont
  {Ngampitipan}}, \ and\ \bibinfo {author} {\bibfnamefont {M.}~\bibnamefont
  {Visser}},\ }\href {\doibase 10.1103/PhysRevD.88.041502} {\bibfield
  {journal} {\bibinfo  {journal} {Phys. Rev. D}\ }\textbf {\bibinfo {volume}
  {88}},\ \bibinfo {pages} {041502} (\bibinfo {year} {2013})}\BibitemShut
  {NoStop}%
\end{thebibliography}%

\bibliographystyle{apsrev4-1}

\end{document}